\documentclass[
reprint,
superscriptaddress,
 amsmath,amssymb,
 aps,
 prl
]{revtex4-2}

\usepackage{graphicx} 
\usepackage{dcolumn} 
\usepackage{bm} 
\usepackage{verbatim} 
\usepackage{xcolor} 
\usepackage{lipsum} 
\usepackage{booktabs} 
\usepackage{multirow} 
\usepackage{hyperref} 
\usepackage{cleveref}
\usepackage[version=4]{mhchem}
\usepackage{bm}
\usepackage{cancel}

\newcommand{\round}[1]{\left( #1 \right)}
\newcommand{\sround}[1]{\left[ #1 \right]}

\newcommand{\Neff}{N_\mathrm{eff}}
\newcommand{\Trh}{T_\mathrm{RH}}

\newcommand{\MeV}{\textrm{MeV}}

\newcommand{\fortepiano}{\texttt{FortEPiaNO}}
\newcommand{\parthenope}{\texttt{PArthENoPE}}

\begin{document}

\preprint{APS/123-QED}

\title{Current constraints on cosmological scenarios with very low reheating temperatures}%

\author{Nicola Barbieri}
\affiliation{Dipartimento di Fisica e Scienze della Terra, Università degli Studi di Ferrara, Via G. Saragat 1, I-44122 Ferrara, Italy}
\affiliation{Istituto Nazionale di Fisica Nucleare (INFN), Sezione di Ferrara, Via G. Saragat 1, I-44122 Ferrara, Italy}
\author{Thejs Brinckmann}
\affiliation{Dipartimento di Fisica e Scienze della Terra, Università degli Studi di Ferrara, Via G. Saragat 1, I-44122 Ferrara, Italy}
\affiliation{Istituto Nazionale di Fisica Nucleare (INFN), Sezione di Ferrara, Via G. Saragat 1, I-44122 Ferrara, Italy}
\author{Stefano Gariazzo}
\affiliation{Department of Physics, University of Turin, via P.\ Giuria 1, 10125 Turin (TO), Italy \looseness=-1}
\affiliation{Istituto Nazionale di Fisica Nucleare (INFN), Sezione di Torino, via P.\ Giuria 1, 10125 Turin (TO), Italy}
\affiliation{Institut de Física Corpuscular (CSIC-Universitat de València), Parc Científic UV, C/ Catedrático José Beltrán 2, E-46980 Paterna (Valencia), Spain}
\author{Massimiliano Lattanzi}
\affiliation{Istituto Nazionale di Fisica Nucleare (INFN), Sezione di Ferrara, Via G. Saragat 1, I-44122 Ferrara, Italy}
\author{Sergio Pastor}
\affiliation{Institut de Física Corpuscular (CSIC-Universitat de València), Parc Científic UV, C/ Catedrático José Beltrán 2, E-46980 Paterna (Valencia), Spain}
\author{Ofelia Pisanti}
\affiliation{Dipartimento di Fisica ``Ettore Pancini'', Università degli studi di Napoli ``Federico II'', Complesso Univ. Monte S. Angelo, I-80126 Napoli, Italy}
\affiliation{INFN - Sezione di Napoli, Complesso Univ. Monte S. Angelo, I-80126 Napoli, Italy}

\date{\today}

\begin{abstract}
We present a comprehensive analysis of the effects of models with very low reheating scenarios ($T_\text{RH} \sim \mathcal{O}(\text{MeV})$) on the cosmological observables and derive corresponding bounds on the reheating temperature. With respect to previous work, our study includes a more precise computation of neutrino distribution functions, leveraging the latest datasets from cosmological surveys.
We perform a joint analysis that combines constraints from Big Bang Nucleosynthesis, the Cosmic Microwave Background, and galaxy surveys, alongside separate investigations of these datasets, carefully assessing the impact of different choices of priors.
At the $95\%$ confidence level, we establish a lower bound on the reheating temperature of $T_\text{RH} > 5.96 \; \text{MeV} $, representing the most stringent constraint to date.
\end{abstract}


\maketitle

\paragraph*{Introduction}

In the standard cosmological picture of the Universe, the expansion rate at early epochs is driven by relativistic particles. This radiation-dominated era arises from the decay products of a massive particle, in a process called reheating. It is usually assumed that this occurred at very large temperatures, after primordial inflation. However, in non-standard cosmological models, long-lived massive particles other than the inflaton might exist, and be responsible for one or more reheating episodes at later times.

In a so-called very low reheating scenario, the start of the radiation-dominated era is significantly delayed with respect to the standard case, down to cosmic temperatures $\sim{\cal O}(1)$ MeV. Since weak processes involving neutrinos are only effective above 1 MeV, the thermalization of the neutrino background could be incomplete due to the lack of interactions. The energy spectra of neutrinos are thus depleted, modifying their contribution to the energy density of radiation and their impact on primordial nucleosynthesis, see e.g.\ \cite{Kawasaki:1999na,Kawasaki:2000en,Hannestad:2004px,Ichikawa:2005vw,Ichikawa:2006vm,DeBernardis:2008zz,deSalas:2015glj,Hasegawa:2019jsa}. 
Interestingly enough, the combination of the latest Planck \cite{Planck:2019nip,Carron:2022eyg}, ACT \cite{ACT:2025fju, ACT:2025tim} and SPT \cite{SPT-3G:2024atg,Camphuis:2025uvg} datasets yields a mild ($\simeq 2\sigma$) preference for a low value of the radiation energy density, a key signature of low-reheating scenarios.

In this letter, we present a complete analysis of the impact of very low reheating temperatures on cosmological observables. With respect to previous studies, we have improved the computation of both the neutrino distribution functions during the decoupling epoch and the primordial abundances of light elements, using modified versions of the \fortepiano\ \cite{Gariazzo:2019gyi,Bennett:2020zkv} and \parthenope\ \cite{Gariazzo:2021iiu} codes, respectively, as well as refining the statistical analysis of data from Cosmic Microwave Background (CMB) observations and galaxy surveys.


\paragraph*{Production of neutrinos in low reheating scenarios} \label{sec:model}

Our non-standard cosmological scenario is inspired by the reheating phase required to end the inflationary epoch, as described in \cite{deSalas:2015glj}. We assume that initially a massive scalar field $\phi$ dominates the energy density, until it decays into standard degrees of freedom with a rate $\Gamma_\phi$, reheating the primeval plasma and leading to the radiation-dominated Universe. Here, we consider that the scalar decays exclusively into electromagnetic particles (see \cite{Kawasaki:2000en,Hasegawa:2019jsa} for the case of hadronic decay), while neutrinos are populated via weak interactions with charged leptons. Therefore, neutrinos can achieve thermal equilibrium with the rest of the plasma only if the $\phi$ decays occur sufficiently early.
Our choice of considering only electromagnetic decays of the scalar field is conservative, in the sense that when hadronic decays are taken into account, the constraints on the reheating temperature become stronger, see e.g.~\cite{Hasegawa:2019jsa}. Intermediate scenarios are also expected to provide stronger effects on the cosmological observables, and consequently to be more disfavored by data than the case discussed here.
 
The equation for the energy density of the scalar field corresponds to that of a decaying non--relativistic species in an expanding Universe,
\begin{equation} \label{eq:rho_phi}
    \frac{d \rho_\phi}{d t} + \left( 3 H + \Gamma_\phi \right) \rho_\phi = 0 \; ,
\end{equation}
where inverse decays are neglected, and $H$ is the Hubble parameter, which depends on the total cosmological energy density. Although reheating is not an instantaneous process, it is convenient to define a reheating temperature, $T_\mathrm{RH}$, at which it is assumed to be completed. Following \cite{deSalas:2015glj}, we define the reheating temperature as $\Gamma_\phi = 3 H (T_\mathrm{RH})$, assuming that from that moment the Universe is fully dominated by radiation. Thus, the Hubble parameter can be expressed as
\begin{equation} \label{eq:hubble_def}
    H \left( T_\mathrm{RH} \right) = \sqrt{\frac{\rho_\mathrm{rad} \left( T_\mathrm{RH} \right)}{3 M^2_\mathrm{pl}}} = \sqrt{\frac{\pi^2}{90} g_* \left( T_\mathrm{RH} \right)} \frac{T^2_\mathrm{RH}}{M_\mathrm{pl}} \; ,
\end{equation}
where $M^2_\mathrm{pl} \equiv m^2_\mathrm{pl}/8\pi = 2.4 \times 10^{18} \; \mathrm{GeV}$ is the reduced Planck mass and $g_* \left( T \right)$ represents the number of relativistic degrees of freedom at a temperature $T$, which takes the value $g_* \left( T \right) = 10.75$ when only photons, neutrinos, and electrons remain in the plasma (the relativistic particles present in the standard case at MeV temperatures). From \cref{eq:hubble_def}, $T_\mathrm{RH}$ and $\Gamma_\phi$ are related as follows
\begin{equation}
    T_\mathrm{RH} \simeq 0.7 \left( \frac{\Gamma_\phi}{\mathrm{s}^{-1}} \right)^{1/2} \; \mathrm{MeV} \; .
\end{equation}
In the remainder of this letter, we focus on the so-called \textit{very low reheating} scenarios, when $T_\mathrm{RH} < 20 \; \mathrm{MeV}$.

In our model, the particle content of the Universe consists of the electromagnetic components ($\gamma$, $e^\pm$, $\mu^\pm$, in equilibrium with a common temperature $T_\gamma$), neutrinos and the massive scalars. Its evolution in time is found solving simultaneously \cref{eq:rho_phi} and the continuity equation for the total energy density, as well as the quantum kinetic equations of the neutrino distribution functions, $f_{\nu_\alpha}(p,t)$. The latter involves a set of integro-differential Boltzmann equations for the neutrino density matrices, necessary in order to take into account the effects of neutrino interactions and flavor oscillations, both relevant for the range of temperatures of interest. This task has been performed using a modified version of \fortepiano\ \cite{Gariazzo:2019gyi}, a fully momentum-dependent decoupling code that provides the evolution of the neutrino density matrices. 

We refer the reader to \cite{Bennett:2020zkv} for a description of the kinetic equations and technical details concerning the numerical computation. In particular, we assume zero neutrino asymmetry, thus neutrinos and antineutrinos share the same density matrices. For the neutrino mixing parameters, we adopt the best--fit values provided in \cite{deSalas:2020pgw} fixing, for simplicity, the CP--violating phase to zero as in~\cite{Gariazzo:2019gyi, Akita:2020szl, Bennett:2020zkv, Froustey:2020mcq, deSalas:2016ztq, Mangano:2001iu}.

\begin{figure}
    \centering
    \includegraphics{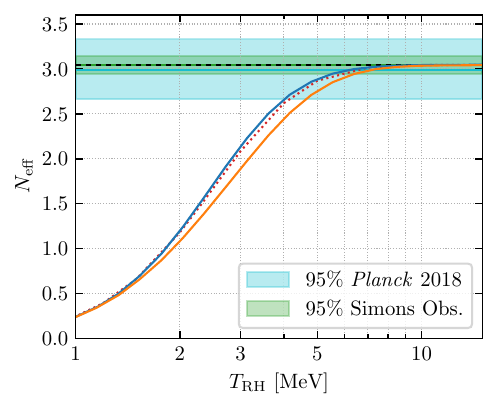}
    \caption{Final neutrino energy density expressed in terms of $N_\mathrm{eff}$, as a function of the reheating temperature. The horizontal line indicates the standard value, $N_\mathrm{eff}=3.044$. Solid lines indicate our new results with (blue) and without (orange) flavor neutrino oscillations. For comparison, the results of the analysis of \cite{deSalas:2015glj} are shown with a dotted red line. Filled regions indicate the present bounds from Planck \cite{Planck:2019nip} and the future sensitivity from the Simons observatory \cite{Ade:2018sbj} on $N_\mathrm{eff}$ at 95\% CL.}
\label{fig:trh_to_neff}
\end{figure}

The incomplete thermalization of neutrinos for MeV reheating temperatures is presented in \cref{fig:trh_to_neff}. For each value of $\Trh$, we show the final contribution of neutrinos to the radiation energy density, expressed in terms of $N_\text{eff}$, the effective number of neutrinos that quantifies the cosmological neutrino-to-photon energy densities,
\begin{equation}\label{eq:neff_def}
    N_\text{eff} = \frac{8}{7}\left (\frac{11}{4}\right)^{4/3} \frac{\rho_\nu}{\rho_\gamma} 
\end{equation}
if there are no other relativistic particles in the primeval plasma. In very low reheating scenarios, when $T_\mathrm{RH} \lesssim 8$ MeV, the neutrino contribution to the relativistic energy can be significantly reduced.

The output of the \fortepiano\ code provides the neutrino distribution functions in the flavor basis, which is convenient for studying neutrino decoupling when weak interactions are still effective. However, after decoupling neutrinos propagate as mass eigenstates, making the physical basis more appropriate for writing the Boltzmann equations in cosmological perturbation theory. The distribution functions in the flavor basis, $f_{\nu_\alpha}\round{p}$ ($\alpha = e$, $\mu$, $\tau$), and in the mass basis, $f_{\nu_i}\round{p}$ ($i = 1$, $2$, $3$), after decoupling are related by
\begin{equation}
    f_{\nu_i} \round{p} = \sum_{\alpha = e, \mu, \tau} \left\lvert U_{\alpha i} \right\rvert^2 f_{\nu_\alpha} \round{p} \; ,
\end{equation}
where $U$ is the neutrino mixing matrix, with all parameters fixed to their best--fit values from~\cite{deSalas:2020pgw}.

\paragraph*{Primordial nucleosynthesis}

Neutrinos influence the production of primordial light element yields during Big Bang Nucleosynthesis (BBN) through two key mechanisms. First, the energy density of all neutrino states contributes to both the Hubble expansion rate and the continuity equation for the radiation energy density. Second, the energy distribution of electron neutrinos directly affects the charged-current weak rates that govern neutron--proton chemical equilibrium. Note that while the first effect is connected to the value of $N_\text{eff}$, the second one depends on the detailed dependence of the neutrino distributions from the comoving momentum; in particular, the two contributions have an opposite behaviour as a function of $T_\mathrm{RH}$, as described in more detail in the Supplemental Material.

In order to obtain the BBN bounds on very low reheating scenarios, we modified the latest version of the \parthenope\ code \cite{Gariazzo:2021iiu} in such a way to implement the neutrino energy density and the electron neutrino distribution calculated by \fortepiano\ \cite{Gariazzo:2019gyi} at a given value of the reheating temperature. With respect to the analysis in \cite{deSalas:2015glj}, where a naive modification of weak rates was implemented in \parthenope, here we apply radiative corrections to the modified Born rates consistently throughout the entire BBN evolution. In addition, we use the neutron lifetime value $\tau_n = \round{879.4 \pm 0.6} \; \text{s}$ \cite{ParticleDataGroup:2022pth}. The results of the \parthenope\ runs show that deuterium and helium yields have an opposite behaviour as functions of $T_\mathrm{RH}$, with deuterium (helium) an increasing (decreasing) function of $T_\mathrm{RH}$. Figures and additional details can be found in the Supplemental Material. 

We performed a BBN likelihood analysis by varying the input parameters of our model, $T_\mathrm{RH}$ and the present value of the baryon density $\omega_\mathrm{b}=\Omega_\mathrm{b} h^2$, where the present-day Hubble parameter is $H_0 = 100h$ km s$^{-1}$ Mpc$^{-1}$. We consider the most recent astrophysical measurements of the primordial abundances from PDG 2022 \cite{ParticleDataGroup:2022pth}: 
$\text{D/H} = (2.547 \pm 0.025) \times 10^{-5}$ for deuterium and $Y_p = 0.245 \pm 0.003$ for helium-4. We also performed the analysis using the helium determination from the EMPRESS survey \cite{Matsumoto:2022tlr}, $Y_p=0.2370\pm0.0034$, about 3$\sigma$ lower than the standard BBN one, obtaining a preference for higher values of the reheating temperature (see the Supplemental Material).

\begin{figure}
    \centering
    \includegraphics{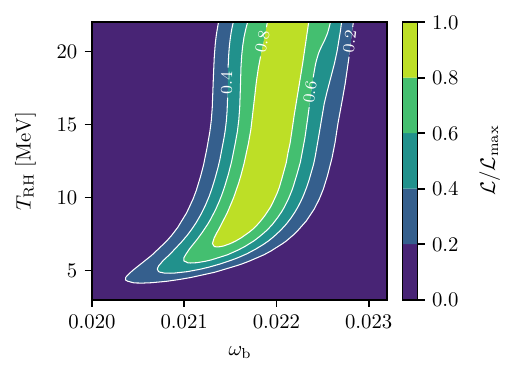}
    \caption{Contour plots for the pure BBN likelihood function (arbitrary units) in the plane $T_\mathrm{RH}-\omega_\mathrm{b}$.}
    \label{fig:BBN_like}
\end{figure}

The contour lines of the likelihood function (in arbitrary units) in the $T_\mathrm{RH}-\omega_\mathrm{b}$ plane are shown in \cref{fig:BBN_like}. The likelihood tends to a constant value for larger values of the reheating temperature, which are equivalent from the BBN perspective. In contrast, it decreases to zero around $T_\mathrm{RH} \sim 3 \; \mathrm{MeV}$, where the predicted primordial abundances are no longer compatible with observations. Adding a prior on $\omega_\mathrm{b}$ from the Planck results narrows the parameter space by cutting off the left and right tails of the likelihood function. However, this prior does not alter the overall trend or conclusions. The results also confirm that BBN data alone cannot constrain the region of larger $T_\mathrm{RH}$, as expected since the production of neutrinos is similar to the standard case.

The correlation between $T_\mathrm{RH}$ and $N_\mathrm{eff}$ presented in \cref{fig:trh_to_neff} can be used for determining a bound on the reheating temperature, once observed that $T_\mathrm{RH}\geq 3$ MeV corresponds to $N_\mathrm{eff}\in\sround{2.14, 3.044}$. In particular, values of $\Neff$ exceeding the upper bound are not permitted in our model, reflecting the theoretical prediction in the standard $\Lambda\text{CDM}$ model \cite{Akita:2020szl,Froustey:2020mcq,Bennett:2020zkv} for very large $T_\mathrm{RH}$. By marginalizing on $\omega_\mathrm{b}$ it is possible to determine the 95\% CL bound on $T_\mathrm{RH}$:
\begin{equation} \label{eq:trh_limit_bbn}
    T_\mathrm{RH} > 3.67 \; \mathrm{MeV} \quad \text{(BBN only)} \; .
\end{equation}

\paragraph*{CMB and galaxy surveys}

Low reheating scenarios are also constrained by CMB and galaxy surveys data. In order to compute the CMB angular power spectra in models with a low $T_\mathrm{RH}$, we modified the Boltzmann solver \texttt{CLASS}~\cite{Lesgourgues:2011rh, Blas:2011rf} and the MCMC sampler \texttt{MontePython}~\cite{Audren:2012wb, Brinckmann:2018cvx}. These modifications allow for an arbitrary neutrino energy spectrum, which can be different for each neutrino mass eigenstate. We compute neutrino spectra on a grid of $T_\mathrm{RH}$ values using \fortepiano, and interpolate these to obtain spectra for arbitrary $T_\mathrm{RH}$. Low reheating models are thus described as extensions of the standard cosmological model (or $\Lambda \mathrm{CDM}$ model) with one additional parameter: the reheating temperature, $T_\mathrm{RH}$.

The free parameters in our MCMC analysis are $\left\lbrace \omega_\mathrm{b}, \omega_\mathrm{c}, 100\theta_\mathrm{s}, \log \round{10^{10} A_\mathrm{s}}, n_\mathrm{s}, \tau_\mathrm{reio}, T_\mathrm{RH} \right\rbrace$, where $\omega_\mathrm{b} \equiv \Omega_\mathrm{b}h^2$ and $\omega_\mathrm{c} \equiv \Omega_\mathrm{c}h^2$ represent the present--day density parameters of baryons and cold dark matter, respectively. The parameter $\theta_\mathrm{s}$ denotes the angular size of the sound horizon at recombination, $A_\mathrm{s}$ is the initial super-horizon amplitude of curvature perturbations (evaluated at the pivot scale $k_{*} = 0.05 \; \mathrm{Mpc}^{-1}$), $n_\mathrm{s}$ is the primordial scalar spectral index, and $\tau_\mathrm{reio}$ is the reionization optical depth. We assume a flat Universe with purely adiabatic scalar primordial perturbations, described by a power--law spectrum. Neutrinos were assumed to be massive with degenerate masses, with the sum of their masses set to the minimum allowed in the normal--ordering, $\sum m_\nu = 0.06 \; \mathrm{eV}$. We checked that our inferences on $T_\mathrm{RH}$ are basically unchanged if this assumption is relaxed. We have also checked that, when both $T_\mathrm{RH}$ and neutrino masses are left free to vary, we obtain the same constraints on $\sum m_\nu$ as in the standard $\Lambda \text{CDM} + \sum m_\nu$ scenarios. Finally, we assume BBN consistency, i.e. the helium fraction $Y_p$ - which impacts the pattern of CMB anisotropies - is computed consistently from the values of the baryon density and reheating temperature from \parthenope. A deeper discussion of both the role of neutrino masses and BBN consistency can be found in the Supplemental Material.

All parameters of the baseline $\Lambda\mathrm{CDM}$ model were sampled from flat prior distributions. 
The reheating temperature was instead sampled from a non-linear prior explicitly constructed to ensure a uniform sampling on $N_\mathrm{eff} \in \sround{2.14, 3.044}$ ($N_\mathrm{eff}$ sampling strategy), which is the range corresponding to $T_\mathrm{RH} \in [ 3, \infty ) \, \mathrm{MeV}$ in our model. All results presented in the main body of this letter are based on this reference prior. Our choice is dictated by the fact that a flat, unbounded prior \footnote{Such a prior would be improper; however, improper priors can be used in Bayesian analyses as long as they yield a proper posterior.} on $T_\mathrm{RH}$ would yield an improper (i.e. enclosing infinite probability mass) posterior for the same parameter, since the likelihood plateaus to a constant, non zero value for sufficiently large values of $T_\mathrm{RH}$. A proper posterior can be obtained by arbitrarily cutting the prior at ``large'' values of $T_\mathrm{RH}$. This was the strategy used in ~\cite{deSalas:2015glj}. The downside of this approach is that the \emph{a posteriori} lower bounds on $T_\mathrm{RH}$ will depend on the arbitrarily chosen \emph{a priori} upper bound. In fact, we have explicitly checked that a $T_\mathrm{RH}$ flat prior would be more informative, in the sense of information theory, than a flat $\Neff$ prior, at least for the actual data realization we are considering here.
We have in any case performed our analysis also considering a flat prior on $T_\mathrm{RH} \in \sround{3, 10} \; \mathrm{MeV}$ ($T_\mathrm{RH}$ sampling strategy), and will briefly comment differences with the flat $\Neff$ prior in the following. A more detailed comparison of the two sampling strategies can be found in the Supplemental Material.

The primary impact of a very low $\Trh$ on the CMB spectra closely resembles that of a cosmological model with $N_\mathrm{eff}<3.044$, and a value of $Y_p$ larger than the standard BBN prediction. Even if subdominant with respect to $\Neff$, the change in $Y_p$ cannot be neglected for a proper analysis of current data. Moreover, the incomplete thermalization also introduces distortions in the neutrino spectra, thus in principle the changes in $\Neff$ and $Y_p$ do not completely capture the effect of a low reheating temperature. Recent results \cite{Alvey:2021sji} suggest that CMB data, both current and next--generation, are unlikely to directly detect specific features in the neutrino distribution function. We further explore the phenomenology of low reheating temperatures, and the role of these effects on CMB spectra, in the Supplemental material.
We choose however to implement all these effects, including the full form of the neutrino spectra, in our CMB analysis, to capture all associated phenomenology with the highest possible accuracy. This will also ensure consistency when performing the joint CMB+BBN analysis, since BBN is more sensitive to spectral distortions.

To compute our parameter constraints, we used the Planck Legacy 2018 CMB temperature, polarization and lensing data and likelihoods, publicly released by the Planck collaboration~\cite{Planck:2018lbu, Planck:2018vyg, Planck:2019nip}. We also incorporated geometric information from measurements of the baryon acoustic oscillations (BAO), based on recent results from the DESI collaboration~\cite{DESI:2024mwx}. For BBN, light element abundance measurements from PDG 2022 \cite{ParticleDataGroup:2022pth} are used. Both DESI and BBN data were included as a Gaussian likelihood. We label the combined data without BBN as ``Planck+lensing+DESI''. When including measurements of light element abundances, we refer to the resulting dataset as ``BBN+Planck+lensing+DESI''. 

We repeated the analysis replacing DESI with BAO measurements from of 6dFGS/SDSS/BOSS/eBOSS~\cite{Beutler:2011hx, Ross:2014qpa, eBOSS:2020qek, eBOSS:2020fvk, eBOSS:2020lta, eBOSS:2020hur, eBOSS:2020gbb, eBOSS:2020uxp, eBOSS:2020tmo,BOSS:2016wmc,eBOSS:2020yzd}. Given that only minor differences are found between the constraining power of the two datasets, we quote here the results with DESI BAO, as the more constraining combination; the results from the other BAO datasets are reported in the Supplemental Material. 

In order to validate our setup, we performed a run with $\Trh$ fixed at $25 \; \mathrm{MeV}$. As shown in \cref{fig:trh_to_neff}, for $T_\mathrm{RH} \gtrsim 10 \; \mathrm{MeV}$, reheating occurs early enough to ensure that all neutrinos reach thermal equilibrium, effectively recovering the standard $\Lambda\mathrm{CDM}$ scenario. Following this test, we proceeded with MCMC runs based on the extended $\Lambda\mathrm{CDM}+T_\mathrm{RH}$ cosmological model.

\begin{figure*}[t]
    \centering
    \includegraphics[width=.99\textwidth]{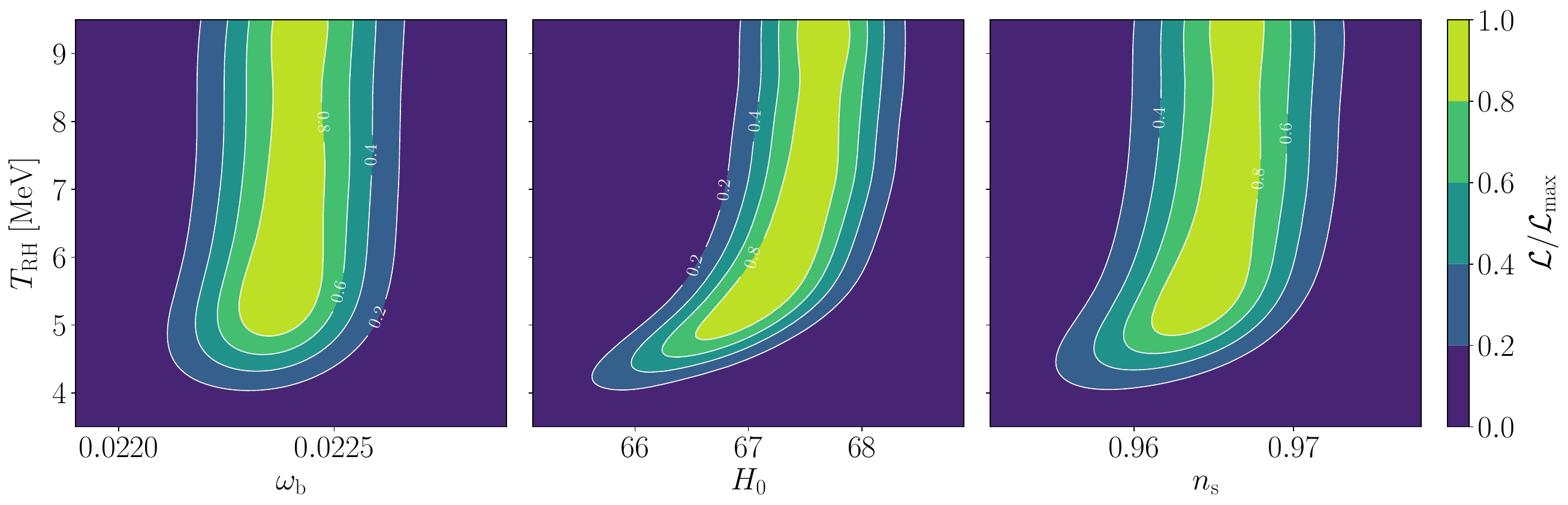}
    \caption{Contour plots for the Planck+lensing+DESI marginalized likelihood function (arbitrary units). The first panel shows the $T_\mathrm{RH}-\omega_b$ plane as a comparison with \cref{fig:BBN_like}. The second and third panels show two key degenerations between parameters of the model.}
    \label{fig:CMB}
\end{figure*}

\paragraph*{Results and discussion}

The marginalized joint likelihood function for the Planck+lensing+DESI dataset is displayed in \cref{fig:CMB}. Across all three panels, a common behavior is evident: for high reheating temperatures, the likelihood plateaus to a constant value, whereas it drops to zero around $T_\mathrm{RH} \sim 3 \; \text{MeV}$, similar to the behavior observed in BBN case. Moreover, in the second panel, a degeneracy with the $H_0$ parameter can be noted for low values of $T_\mathrm{RH}$. This is a reflection of the well-known $H_0 - N_\mathrm{eff}$ correlation. The direction of the correlation, together with the fact that $N_\mathrm{eff}<3.044$,
imply that low reheating models do not provide a resolution to the Hubble tension. Finally, in the third panel, it can be observed that low reheating temperatures show a slight preference for low values of $n_s$, which may have implications for the capability of cosmological data to evaluate the viability of inflationary models. A more detailed discussion on this topic can be found, for instance, in \cite{Gerbino:2016sgw}.

We obtain the following $95\%$ lower bound on the reheating temperature from Planck+lensing+DESI:
\begin{equation} \label{eq:trh_limit_neff}
    T_\mathrm{RH} > 3.79 \; \mathrm{MeV} \quad \text{(Planck+lensing+DESI)} \; ,
\end{equation}
and the corresponding bound on $N_\mathrm{eff}$:
\begin{equation} \label{eq:neff_limit_neff}
    N_\mathrm{eff} > 2.58 \quad \text{(Planck+lensing+DESI)} \; .
\end{equation}
We find tighter bounds when considering instead a flat prior on $T_\mathrm{RH} \in \sround{3, 10} \; \MeV$. This is in part due to the fact that this prior gives larger statistical weight to models with $N_\mathrm{eff}$ close to $3.044$, thus being less tolerant to deviations from the standard picture than our reference flat $N_\mathrm{eff}$ prior.

It is important to correctly implement BBN consistency.
As noted above, a higher $Y_p$ is associated to low $T_\mathrm{RH}$. This increases the photon diffusion length, partly counterbalancing the effect of a low $N_\mathrm{eff}$. We have explicitly checked that (incorrectly) disregarding BBN consistency would lead to slightly stronger constraints than Eq.~\ref{eq:trh_limit_neff}. This is less evident when using the flat $T_\mathrm{RH}$ prior, because this concentrates most of the probability around $N_\mathrm{eff}=3.044$.

Finally, we carried out a series of runs with a combined dataset that included CMB and galaxy survey data together with the BBN measurements of primordial abundances. The inclusion of BBN data significantly enhances the constraining power of the dataset. Our final $95\%$ CL lower limits on the reheating temperature and $N_\mathrm{eff}$ are 
\begin{align}
&T_\mathrm{RH} > 5.96 \; \mathrm{MeV} \; \text{(BBN+Planck+lensing+DESI)} \;,\\[0.2cm]
&N_\mathrm{eff} > 2.98 \; \text{(BBN+Planck+lensing+DESI)} \; .
\end{align}
This is the most stringent bound to date on the reheating temperature.

\paragraph*{Conclusions}

The onset of the radiation-dominated era of the Universe is delayed in low reheating scenarios with respect to the standard cosmological model. For values of $\Trh$ as low as $\mathcal{O}(\text{MeV})$, neutrino production can be significantly reduced with important implications on BBN and later cosmological phases. Here we have carried out a full analysis of very low reheating models improving on the results of previous works
\cite{Kawasaki:1999na,Kawasaki:2000en,Hannestad:2004px,Ichikawa:2005vw,Ichikawa:2006vm,DeBernardis:2008zz,deSalas:2015glj,Hasegawa:2019jsa} with an improved computation of the evolution of neutrino spectra through the decoupling process and a refined calculation of the production of light elements at BBN. 

We carried out the statistical study with a new approach, including the latest available datasets for all the considered probes. We find the most stringent  bound to date on the reheating temperature, $T_\mathrm{RH} > 5.96 \; \mathrm{MeV}$ ($95\%$ CL) with the full dataset including Planck, lensing, BAO and measurements of primordial abundances. Such lower limit is relevant for any theoretical cosmological model with a very low reheating temperature.

Our results are also important for scenarios with additional relativistic particles, whose contribution to the radiation energy density could be reduced for very low values of $\Trh$. In particular, when light sterile neutrinos exist with non-zero mixing with the active states, a case where low reheating temperatures could relax the tight cosmological bounds in the standard $\Lambda \mathrm{CDM}$ model \cite{Gelmini:2004ah,Yaguna:2007wi,Gelmini:2008fq,Abazajian:2017tcc,Gelmini:2019wfp,Hasegawa:2020ctq}.

Given the role played by $\Neff$ in models with a low-reheating temperature, we can expect that next-generation CMB experiments will improve constraints on these scenarios. A preliminary analysis, described in the supplemental material, shows that indeed we can expect that CMB observations alone will be as constraining as the combination of current CMB and primordial abundances data.

\begin{acknowledgments}
\paragraph*{Acknowledgements}
    We acknowledge the use of GetDist \cite{Lewis:2019xzd} software package, and the use of computing facilities provided by the INFN theory group (I.S. InDark) at CINECA. 
    N.B., S.G.\ and S.P.\ would like to thank the Galileo Galilei Institute for Theoretical Physics for the hospitality and the INFN for partial support during the completion of this work. 
    O.P.\ is supported by Ministero dell’Università e della Ricerca (MUR), PRIN2022 program (Grant PANTHEON 2022E2J4RK) Italy.
    O.P.\ and S.G.\ are supported by the Research grant TAsP (Theoretical Astroparticle Physics) funded by Istituto Nazionale di Fisica Nucleare (INFN).
    S.G.\ also acknowledges financial support through the Ram\'on y Cajal contract RYC2023-044611-I funded by MICIU/AEI/10.13039/501100011033 and FSE+.
    S.P.\ is supported by the Spanish grants PID2023-147306NB-I00 and CEX2023-001292-S (MCIU/AEI/10.13039/501100011033), as well as CIPROM/2021/054 (Generalitat Valenciana).
    T.B.\ was supported by ICSC -- Centro Nazionale di Ricerca in High Performance Computing, Big Data and Quantum Computing, funded by European Union -- NextGenerationEU. M.L.\ acknowledges the financial support from the INFN InDark initiative and from the COSMOS network
    (\url{www.cosmosnet.it}) through the ASI (Italian Space Agency) Grants 2016-24-H.0 and 2016-24-H.1-2018, as well as 2020-9-HH.0 (participation in LiteBIRD, phase A). M.L.\ is partially funded by the European Union (ERC, RELiCS, project number 101116027).
\end{acknowledgments}

\bibliography{main}

\clearpage
\onecolumngrid

\begin{center}
    \textbf{\large Supplemental Material for\\``Current constraints on cosmological scenarios with very low reheating temperatures''}\\[5pt]

    \vspace{0.1cm}
    \begin{quote}
        {\small In this Supplemental Material, we provide additional figures and details on some points discussed in the main text. Specifically, we discuss how the abundances of light elements depend on the reheating temperature, the comparison between different sampling strategies and the impact of neutrino masses on our analysis. Finally, we provide a summary discussion of all the reheating temperature constraints derived with different procedures and data set combinations within this work.}\\[10pt]
    \end{quote}
\end{center}

\makeatletter
  \setcounter{figure}{0}
  \renewcommand{\thefigure}{S\arabic{figure}}
  \renewcommand{\theHfigure}{S\arabic{figure}}

  \setcounter{table}{0}
  \renewcommand{\thetable}{S\arabic{table}}
  \renewcommand{\theHtable}{S\arabic{table}}

  \setcounter{equation}{0}
  \renewcommand{\theequation}{S\arabic{equation}}
  \renewcommand{\theHequation}{S\arabic{equation}}

  \setcounter{section}{0}
  \renewcommand{\thesection}{S\Roman{section}}
  \renewcommand{\theHsection}{S\Roman{section}}
\makeatother

\vspace{0cm}
\thispagestyle{empty}
\pagenumbering{arabic}

\section{Light elements abundances at low reheating temperature}

\begin{figure*}[h!]
        \centering
        \includegraphics[scale=0.9]{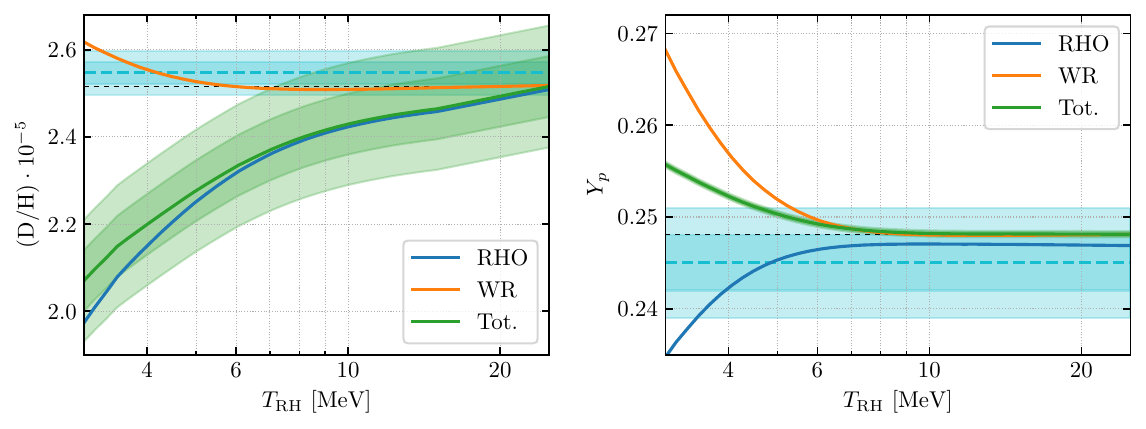}
        \caption{Deuterium (left) and helium (right) abundances as a function of the reheating temperature for $\Omega_b h^2=0.02242$ (see the text for a detailed discussion). Shaded cyan regions correspond to the $68\%$ and $95\%$ CL regions recommended in \cite{ParticleDataGroup:2022pth}, while shaded green regions correspond to $68\%$ and $95\%$ theoretical errors.}
        \label{fig:bbn_trh}
    \end{figure*}

    In \cref{fig:bbn_trh} we present the $\ce{D}/\ce{H}$ and $Y_p$ abundances as functions of the reheating temperature. In addition to the total final abundances (shown by the green solid line), we separately show the two individual contributions cited in the main text: RHO (blue lines), due to the energy density of all neutrino states in both the Hubble expansion rate and the continuity equation for the radiation energy density, and WR (orange lines), corresponding to the energy distribution of electron neutrinos in the charged-current weak rates.
    
    These two different contributions have an opposite behaviour as a function of $T_\mathrm{RH}$. Concerning the RHO contribution, both deuterium and helium yields increase as functions of $T_\mathrm{RH}$. This is because, at low reheating temperatures, the production of neutrinos is less efficient, leading to a lower expansion rate and fewer neutrons available for the synthesis of light elements. In contrast, at low reheating temperatures, variations in the electron neutrino distribution result in a decrease in weak rates (WR contribution), causing an earlier decoupling of the processes responsible for proton--neutron interconversion. This leads to a higher freeze-out value of the $n/p$ ratio, which increases the final value of $Y_p$, and to a lesser extent, the deuterium abundance. 

    As already stressed, only the RHO contribution can be traced back to a change in the value of $N_\text{eff}$, while the WR one is produced by the distortions in the neutrino spectra, which move further away from the Fermi-Dirac form as $T_\mathrm{RH}$ decreases. This implies that one cannot rely only on an ``integrated quantity" like $N_\text{eff}$ for estimating the behaviour of the abundances with $T_\mathrm{RH}$: it is the combination of the two competing effects that explains the net increase of deuterium towards high $T_\mathrm{RH}$, due to the dominant contribution of RHO, in contrast with the rise of $Y_p$ at low $T_\mathrm{RH}$, where the WR contribution is larger.

    In the analyses found in literature several combinations of deuterium and helium abundances are commonly used (see for example ref.~\cite{Schoneberg:2024ifp} for an update). In particular, different helium determinations are more or less in agreement with the exception of the recent measurement from the EMPRESS survey \cite{Matsumoto:2022tlr}, $Y_p=0.2370\pm0.0034$, about 3$\sigma$ lower than the standard BBN one. As evident from the right plot in \cref{fig:bbn_trh}, a lower helium abundance goes in the direction of a preference for higher values of the reheating temperature. In fact, by repeating the BBN-only analysis with the EMPRESS value one obtains contour plots shifted towards higher values of $T_\mathrm{RH}$ and $\omega_\mathrm{b}$. Then, the choice made in this paper for $Y_p$ represents a conservative one, other than being in agreement with the general feeling that a lower value of the helium abundance is an 'outlier' until further evidence. 

\section{CMB spectra at low reheating temperature \label{sec:cmbpheno}}

    In this section we want to examine in more detail the effect of a low reheating temperature on the CMB power spectra. The incomplete thermalization of the neutrinos results in a lower neutrino energy density and in nonthermal features in their phase space distribution, i.e., spectral distortions. With respect to the standard instantaneous decoupling scenario, where neutrinos follow a Fermi-Dirac distribution with $T_\nu=(4/11)^{4/3} T_\gamma$, the  neutrino distribution function for low reheating temperatures can in principle be obtained by first suitably rescaling the temperature (or normalization) in order to get the energy density (i.e., $\Neff$) corresponding to a given $\Trh$, and then adding spectral distortions. Both these effects can in principle alter the evolution of cosmological perturbations, the lower value of $\Neff$ acting at both the background and perturbation level, and the spectral distortions acting only at the perturbation level. Although one might expect that the dominant effect is that related to the low density, it is less straightforward to guess the relevance of spectral distortions in the evolution of cosmological perturbations.

    In addition to the above, BBN is also altered in low-reheating scenarios, again because of a combination between the changes in the expansion history induced by the low $\Neff$, and the momentum-dependent changes in the weak rates induced by the spectral distortions, as detailed in the previous section. All this is also relevant for CMB observables, that are sensitive, through the time of recombination, to changes in the $^4\mathrm{He}$ abundance.

    In this section we thus seek to quantify the relative importance for CMB observables of these effects.
    In order to do this, we consider here three different implementations of the calculation of CMB observables in low reheating scenarios. The first implementation, that we take as a reference, uses the neutrino spectra computed from \fortepiano, fully accounting for the lower density and non-thermal distortions, in the Boltzmann hierarchy within \texttt{CLASS}. It further uses, as an input to \texttt{CLASS}, the helium abundance consistently computed from \parthenope\ using those same non-thermal neutrino spectra. This is the same implementation used for the analysis presented in the main text, and we regard it as providing the most accurate prediction of the theoretical CMB power spectra for a given value of $\Trh$. We will refer to this implementation as ``full PSD + $Y_p$'', or simply ``exact''. 
    We further consider two additional implementations, ``$\mathrm{FD}$'' and ``$\mathrm{FD}+Y_p$''. In both, we approximate the neutrino distribution function as Fermi-Dirac, suitably rescaled in order to have the value of $\Neff$ corresponding to a given $\Trh$
    \footnote{This is somehow analogous to what is usually done, in the standard cosmological scenarios, when accounting for a more accurate modeling of neutrino decoupling in the calculation of CMB observables. In fact, it is well known that an accurate treatment of neutrino decoupling yields a larger energy density ($\Neff=3.044)$ than expected for instantaneous decoupling, and spectral distortions. Still, this is usually taken into account, in CMB analysis, by neglecting spectral distortions and using instead a Fermi-Dirac distribution for the neutrinos (like in the instantaneous decoupling scenario), just with slightly higher temperature to produce $\Neff=3.044$. This approximation is sufficient to give the precision of current CMB data.}. In the $\mathrm{FD}$ approximation, the helium abundance is (incorrectly) computed using the standard BBN consistency relation included in \texttt{CLASS}, that only takes into account the changes in the expansion history at BBN time induced by a change in $\Neff$. In the $\mathrm{FD}+Y_p$ implementation the helium abundance is instead computed as in the ``full PSD + $Y_p$'' case, i.e. accounting for both the changes in expansion history and for the effect of neutrino spectral distortions. In other words, the $\mathrm{FD}$ implementation only considers the change in $\Neff$ (both at the BBN and CMB times) but altogether ignores the effect of spectral distortions beyond the change in $\Neff$. The $\mathrm{FD}+Y_p$ implementation instead neglects the distortions for the purpose of the neutrino perturbation equations, but accounts for them in the helium yield that is provided as an input to \texttt{CLASS}. We thus expect $\mathrm{FD}+Y_p$ to provide a better approximation than FD for the CMB observables in a low reheating scenario.

    \begin{figure*}[t]
            \centering
            \includegraphics[scale=0.70]{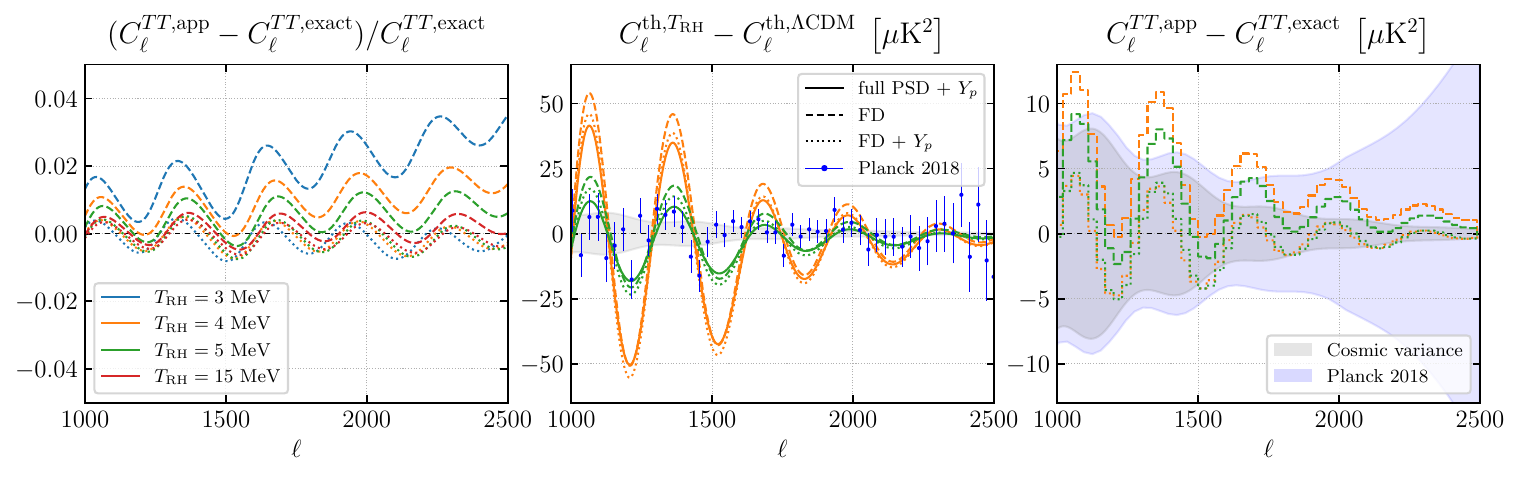}
            \caption{Comparison between the approximated (FD and FD+$Y_p$, dashed and dotted lines respectively, in all plots) and exact (PSD+$Y_p$, solid) computation of CMB temperature power spectra. The approximated implementations use a Fermi-Dirac distribution for the neutrinos inside the Boltzmann code \texttt{CLASS}, while the exact one accounts for the actual form derived from \fortepiano, thus including spectral distortions. In the FD implementation the helium abundance is computed accounting only for changes in the expansion history in the radiation-dominated era. In the FD+$Y_p$ and PSD+$Y_p$ implementations this is instead computed also taking into account the nonthermal neutrino distribution. See text for details. Left panel: Relative difference between spectra computed using the FD or FD+$Y_p$ and the PSD+$Y_p$ implementation, for different values of the reheating temperature. 
            Middle panel: Difference between spectra for different values of the reheating temperature and the Planck 2018 best-fit model, using the three implementations under considerations. Also shown are the Planck 2018 data with the corresponding uncertainty. The grey shaded area shows the uncertainty of a cosmic-variance limited experiment observing 70\% of the sky, with the same multipole binning $\Delta \ell = 30$ as the Planck data.
            Right panel: Difference between spectra computed using the FD or FD+$Y_p$ and the PSD+$Y_p$ implementation, for different values of the reheating temperature. In this plot, the theoretical spectra have been binned following the same scheme as the Planck data. The shaded regions show the observational uncertainty of Planck and of the same cosmic-variance limited experiment as in the middle panel.
            \label{fig:cl_diffs}}
    \end{figure*}
    
    We compare the two approximations with the exact result in \cref{fig:cl_diffs}. In the left panel, we show the relative difference between the approximated and exact temperature power spectra, $\left(C_\ell^{TT,\mathrm{app}}-C_\ell^{TT,\mathrm{exact}}\right)/C_\ell^{TT,\mathrm{exact}}$, for different values of $\Trh$, where ``app'' stands for either ``FD'' or ``FD+$Y_p$'',  We stress that the curves with the same color have the same value of $\Neff$. Using the less accurate approximation, FD, and considering values of $\Trh$ between $3$ and $15\,\MeV$ we find that the relative difference in the spectra (solid lines) follows an oscillatory pattern with an amplitude of up to a few \% at small scales. The shape and sign of the difference is consistent with being mostly due to the difference in the value of $Y_p$. From the results of the previous section, we know that the FD implementation is always underestimating $Y_p$, and consequently the photon damping length $r_d \propto(1-Y_p)^{1/2}$ \cite{Hu:1995fqa,Hou:2011ec}. The final effect on the CMB spectra is to raise and slightly shift the damping tail, consistent with what is seen in \cref{fig:cl_diffs}.
    The difference anyway decreases for an increasing reheating temperature, as expected given that the neutrino distribution approaches a thermal shape in the large $\Trh$ limit. Considering instead the FD+$Y_p$ approximation, its relative accuracy is sub-percent (dashed curves), even for reheating temperature as low as $3\,\MeV$. Part of the difference might be an artefact stemming from the difference in the numerical implementation considered here. However, we can still conclude that the specific effect of the neutrino spectral distortions, and thus the accuracy of the FD+$Y_p$ implementation, is sub-percent or lower.

    In the middle panel of \cref{fig:cl_diffs}, we instead show the absolute difference $C_\ell^{\mathrm{th},\Trh}-C_\ell^{\mathrm{th},\Lambda\mathrm{CDM}}$ between the prediction for low-reheating scenarios with $\Trh = 4$ or $5\,\MeV$, computed using either the exact or approximated implementations, and the Planck $\Lambda$CDM best fit model. In the same figure we also show the Planck binned data, with the corresponding uncertainty, and the cosmic variance corresponding to the Planck best fit model, a sky fraction $f_\mathrm{sky} = 0.7$ and a binning $\Delta\ell=30$ (the same as the Planck data). Focusing for the moment only on the theoretical spectra, the plot shows how most of the difference with respect to $\Lambda$CDM is due to the low neutrino energy density; the residual difference is accounted by the different helium abundance and by the effect of neutrino spectral distortions on the evolution of perturbations, in that order of importance.

    Finally, in the right panel we show $\left(C_\ell^{TT,\mathrm{app}}-C_\ell^{TT,\mathrm{exact}}\right)$, with $\textrm{app} = \textrm{FD or FD}+Y_p$, for two values of the reheating temperature ($\Trh = 4$ or $5$ MeV). These differences are compared to the uncertainty of the Planck data, and to that of a cosmic-variance limited experiment (up to $\ell_\mathrm{max} = 2500)$ observing $70\%$ of the sky. The theoretical  curves and the cosmic variance are binned following the same scheme as the Planck data (bin size $\Delta\ell=30$ and weights proportional to $\ell(\ell+1)$) to ease the comparison.

    We can now give a qualitative, but informed, assessment of the suitability of the above approximations for the purpose of parameter estimation from current and future data. We start by noting how, even for $\Trh = 5\,\MeV$, which is well within our $95\%$ credible interval from dataset combinations including Planck and BAO data (see below), the FD approximation is only accurate at the $\sim 1\%$ level over a wide range of scales (left panel of \cref{fig:cl_diffs}, dashed green curve). Such a level of accuracy is not enough given the precision of current CMB measurements.  This is evident from the right panel of \cref{fig:cl_diffs}, that shows how the distance between the FD and exact computation is tipically comparable to the uncertainties in the Planck data. A similar comparison for the FD+$Y_p$ implementation suggests that this approximation might affect parameter inferences from current data in a small but possibly non negligible way. Similar results are obtained for polarization power spectra, with a reduced significance of the different implementations due to the larger error bars on data points.

    The findings of this section can be summarized as follows. The CMB spectra are sensitive to low values of $\Trh$ mainly through their effect on the neutrino energy density (smaller $\Neff$) and on the $^4\mathrm{He}$ yield (larger $Y_p$), in that order of importance. These effects have both to be taken into account when deriving constraints from CMB data. In particular, since increasing $Y_p$ partially compensates for the effect of a smaller $\Neff$ in the CMB damping tail, we expect that neglecting the impact of low reheating on $^4\mathrm{He}$ (i.e., underestimating $Y_p$) will produce artificially stronger bounds on $\Trh$. The effect of the neutrino spectral distortions on the evolution of perturbations is instead less relevant, and appears to be sub-leading if not at all negligible, at least for current data.
    A precise assessment, also in view of future data, would require a dedicated Monte Carlo run that is however beyond the scope of this section. In any case, the main results presented in the paper always rely on the ``exact'' implementation accounting for the full effect of the neutrino spectral distortions.

\section{Comparison between prior choices}

    In this section we discuss our choice of the prior $\Pi$ on $T_\mathrm{RH}$. Naively, the simpler choice might seem to take a flat prior on the parameter, i.e. $\Pi = \mathrm{const}$ in some (wide) range. Ideally, one might directly take a prior constant over the infinite range $T_\mathrm{RH}\in [ 0,\,\infty)$. This would not be a good probability distribution, since it cannot be normalized to unity, but would still be allowed in a Bayesian analysis as long as the integral of its product with the likelihood ${\mathcal L}$ is finite - i.e. as long as the product (prior)$\times$(likelihood) can be normalized to unity to yield a proper posterior. Alternatively, one might start with a proper, but very wide prior on the parameter, and hope (or better check a posteriori) that parameter inferences do not depend on the prior range. This again requires that the likelihood vanishes quickly enough at the prior edges. This is what usually happens with e.g. the standard $\Lambda$CDM parameters. However, this is unfortunately not the case for $T_\mathrm{RH}$, since the likelihood does not vanish for large values of the parameter. The likelihood instead flattens to a maximum value for values larger than a few MeV. This is because these scenarios are essentially indistinguishable from standard $\Lambda$CDM. Thus an improper prior is not a viable choice, and a proper prior will still make inferences on $T_\mathrm{RH}$ dependent on the prior range. It is easy to see why this is the case. We obtain the posterior ${\mathcal P}$ on $T_\mathrm{RH}$ from Bayes' theorem:
    \begin{equation}
        {\mathcal P}(T_\mathrm{RH}) = \frac{{\mathcal L} (T_\mathrm{RH}) \Pi (T_\mathrm{RH})}{{\mathcal Z}} \, .
    \end{equation}
    Here, ${\mathcal Z}$ is the evidence, and we have suppressed the data in the arguments of the various probability distributions since these are fixed for the present purpose. Also, it should be understood that the likelihood ${\mathcal L} (T_\mathrm{RH})$ is averaged over the prior of the $\Lambda$CDM parameters $\boldsymbol{}{\theta}_{\Lambda\mathrm{CDM}}$, i.e. ${\mathcal L} (T_\mathrm{RH}) \equiv \int {\mathcal L} (\boldsymbol{}{\theta}_{\Lambda\mathrm{CDM}}, T_\mathrm{RH}) \Pi(\boldsymbol{}{\theta}_{\Lambda\mathrm{CDM}}) d\boldsymbol{}{\theta}_{\Lambda\mathrm{CDM}}$. Denoting with $T_\mathrm{RH,\,max}$ the prior upper bound, and performing a one-sided analysis (we will come back to this later), the 95\% Bayesian posterior lower bound $T_\mathrm{RH,\,low}$ is defined through
    \begin{equation}
        \int_{T_\mathrm{RH,\,low}}^{T_\mathrm{RH,\,max}} {\mathcal P}(T_\mathrm{RH}) = 0.95 \, .
    \end{equation}
    If the integrand does not vanish above for large enough $T_\mathrm{RH}$, the value of $T_\mathrm{RH,\,low}$ that yields an integrated probability of 0.95 will depend on ${T_\mathrm{RH,\,max}}$ - our inferences on the smaller value of $T_\mathrm{RH}$ allowed by the data will depend by an arbitrary choice on the prior range. This is a less than ideal situation that we would like to avoid if possible.

    The solution that we propose is to adopt a flat prior on $N_\mathrm{eff}$ instead. In this way, the prior is naturally bounded from above because $T_\mathrm{RH}\to \infty$ is  mapped to $N_\mathrm{eff}=3.044$. Since at the lower edge the posterior is naturally cut by the likelihood (this was obviously true also for the $T_\mathrm{RH}$ prior), the resulting posterior is automatically proper and bounds on $N_\mathrm{eff}$ or $T_\mathrm{RH}$ will be independent of arbitrary choices about the prior range. Another reason to argue in favor of the $N_\mathrm{eff}$ flat prior is that the data, and in particular CMB data, are sensitive to $T_\mathrm{RH}$ through $N_\mathrm{eff}$. In other words, the latter is the quantity that is more directly constrained by the data. We note that the likelihood is to a good approximation Gaussian in $N_\mathrm{eff}$. In these terms, a flat prior on $T_\mathrm{RH}$ appears odd, as it concentrates a lot of the probability mass in a small interval around $N_\mathrm{eff}=3.044$, where the data have no discriminating power (i.e., the likelihood is constant). At the same time, the region where the likelihood varies significantly with respect to its maximum value, i.e,. where the data are informative, occupies a small fraction of the total prior volume. Intuitively, one might therefore think that a uniform prior on $N_\mathrm{eff}$ would allow to maximize the information that comes from the data. This intuitive reasoning can be backed up using information theory-based arguments, as we shall see below.

    In \cref{fig:kullback_leibler}, we show the prior and posterior probability distributions for $T_\mathrm{RH}$ corresponding to the two prior choices, i.e. taking a uniform prior on either $N_\mathrm{eff}$ or $T_\mathrm{RH}$. The blue and green curves are the flat prior on $T_\mathrm{RH}$ and the corresponding posterior obtained from the Planck+lensing+DESI likelihood. Note that, apart from a normalization factor, the green curve is also equal to the likelihood. The orange and red curves are instead the non-uniform $T_\mathrm{RH}$ prior induced by a flat $N_\mathrm{eff}$ prior, and the corresponding posterior.

    \begin{figure*}[t]
        \centering
        \includegraphics[scale=0.9]{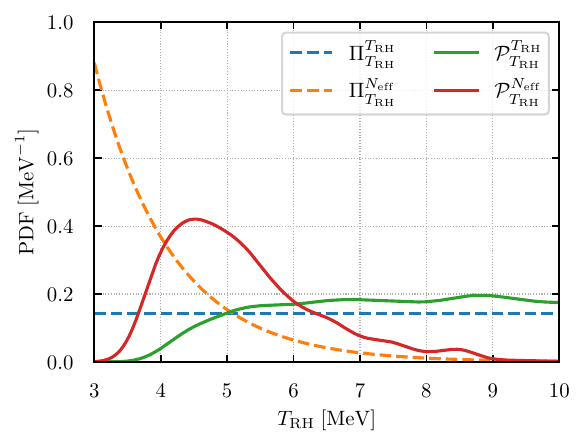}
        \caption{Prior and posterior distributions for $T_\text{RH}$ corresponding to two different prior choices. The blue and green curves are the uniform prior on $T_\mathrm{RH}$ and the corresponding posterior, respectively. 
        The orange and red curves are instead the non-uniform $T_\mathrm{RH}$ prior induced by a flat $N_\mathrm{eff}$ prior, and the corresponding posterior. Solid lines posterior distributions refer to Planck+lensing+DESI data set, while dashed lines refer to Planck+lensing+BOSS/eBOSS data set.}
        \label{fig:kullback_leibler}
    \end{figure*}

    Looking at the orange curve in \cref{fig:kullback_leibler}, one might be bothered that the flat prior on $N_\mathrm{eff}$ is somehow ``more informative'' than the flat $T_\mathrm{RH}$ prior, given that it places more weight to low values of $T_\mathrm{RH}$. This intuition is however not correct, as we show in the following. We can assess how informative a prior is in terms of the Kullback--Leibler (KL) divergence, or relative entropy, between the prior and the posterior for $T_\mathrm{RH}$. This is given by
    \begin{equation} \label{eq:kl_def_text}
        D_\mathrm{KL} \round{\mathcal{P}  \parallel \Pi } = 
            \int d\theta \; \mathcal{P} \round{\theta \mid D} \log \sround{\frac{\mathcal{P} \round{\theta \mid D}}{\Pi \round{\theta}}} \; ,
    \end{equation}
    where $\theta$ is the parameter vector ($\theta = T_\mathrm{RH}$ in this specific case) and $D$ is the data. From a Bayesian perspective, the KL divergence measures the information gain provided by the data in updating the prior distribution, $\Pi$, to the posterior, $\mathcal{P}$. A larger divergence indicates that the data (the likelihood) contribute more significantly to the posterior with respect to the prior. Note that the KL divergence is independent from the particular choice of parametrization.

    We have computed the KL divergence for the prior-posterior pairs in \cref{fig:kullback_leibler} corresponding to the two prior choices under consideration, and obtained
    \begin{align}
        D_\mathrm{KL} = 0.63 \quad (\text{uniform $N_\mathrm{eff}$ prior}) \, , \\
        D_\mathrm{KL} = 0.15 \quad (\text{uniform $T_\mathrm{RH}$ prior}) \,  ,
    \end{align}
    for the Planck+lensing+DESI dataset. Similarly, for Planck+lensing+BOSS/eBOSS: 
    \begin{align}
        D_\mathrm{KL} = 0.43 \quad (\text{uniform $N_\mathrm{eff}$ prior}) \, , \\
        D_\mathrm{KL} = 0.15 \quad (\text{uniform $T_\mathrm{RH}$ prior}) \,  .
    \end{align}

    These results shows that the uniform $N_\mathrm{eff}$ prior is indeed \emph{less} informative (i.e., yields a larger KL divergence) than the uniform $T_\mathrm{RH}$ prior, at least for the particular data realizations considered here. This is another argument in favor of preferring the former to the latter.

    Let us conclude this section by discussing how we report parameter constraints, in particular how we build Bayesian credible intervals for $T_\mathrm{RH}$. There exist infinitely many intervals that enclose a given fraction $\alpha$ (e.g. 95\%) of the total probability. Thus another condition has to be given to fix the interval. A popular choice is to take the highest-probability interval, i.e. such that the probability everywhere inside the interval is larger than the probability everywhere outside. This amount to choosing the shortest interval for a given $\alpha$. One advantage of this choice is that it automatically produces a one-tailed interval if the peak of the posterior is close enough to the posterior boundaries. However, while variable transformations preserve probability mass, they do not preserve volume nor probability density. After a change in parameterization, the 95\% highest-density interval will still be a 95\% interval, but not necessarily the highest-density one. We thus have to make a choice on the parameter used to define the shortest interval. Given the considerations above, we choose to report the 95\% interval that has the highest density in $N_\mathrm{eff}$. This leads us to report a one-sided interval (i.e., a lower limit) on $T_\mathrm{RH}$ even if the posterior for this parameter has a well-definite peak.


\begin{figure*}[t]
        \centering
        \includegraphics[scale=.9]{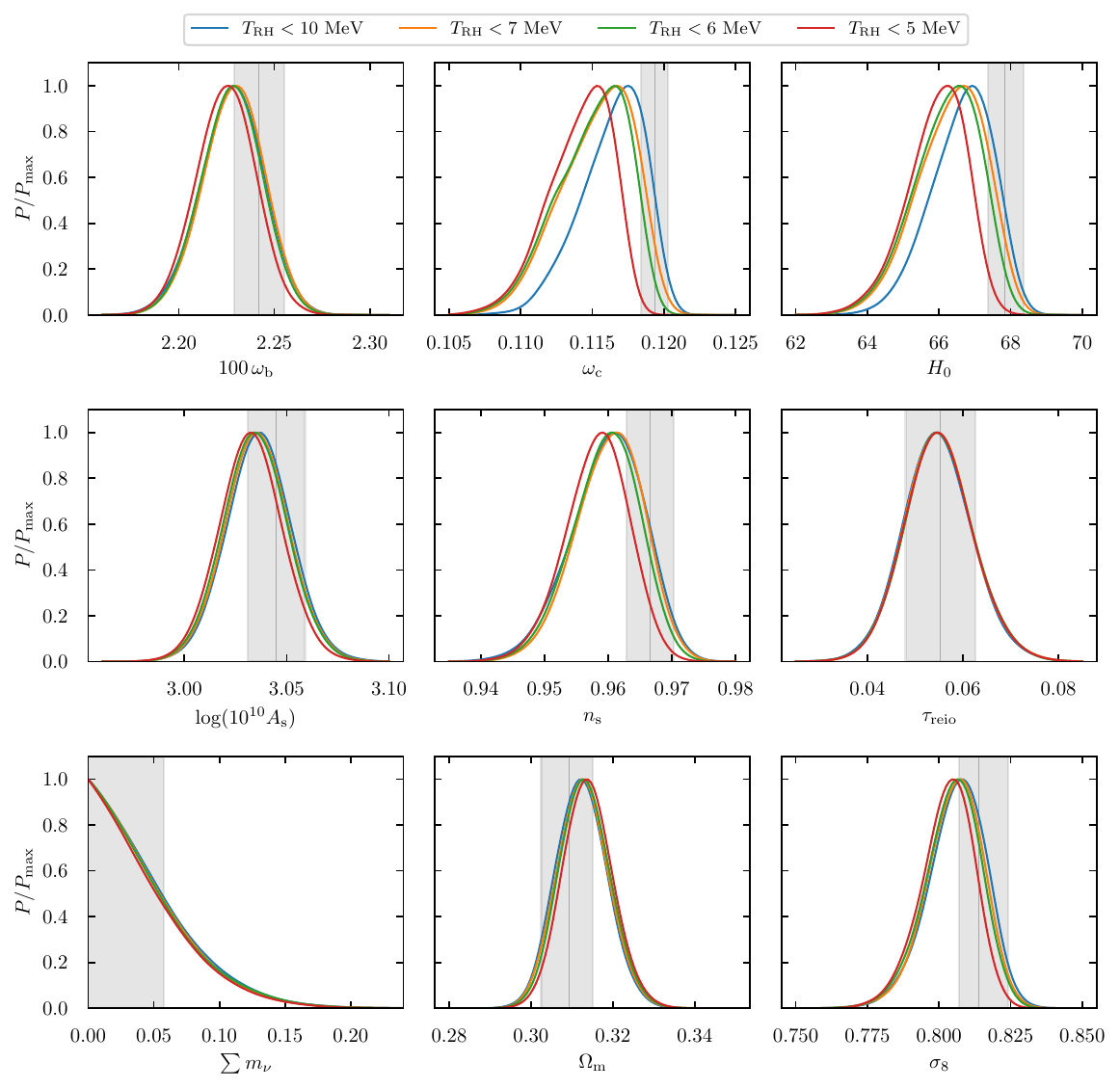}
        \caption{One--dimensional posteriors for selected cosmological parameters in a $\Lambda\text{CDM}+T_\text{RH}+\sum m_\nu$ model from Planck+lensing+BOSS/eBOSS. Each posterior is shown for four different prior cuts on $\Trh$, namely $\Trh < \{5,\,6,\,7,\, 10\}\,\MeV$, in order to highlight parameters variations due to the effect of low reheating temperatures. All the posteriors refer to runs carried out with a flat prior on $N_\text{eff}$. Shaded gray areas represent the Planck 2018 TTTEEE+lensing+BAO constraints on $\Lambda\text{CDM}+\sum m_\nu$ model from \cite{Planck:2018vyg}.}
        \label{fig:params_panel}
\end{figure*}

\newpage
\section{Impact of low-reheating on $\Lambda$CDM and neutrino mass parameters}

    In this section we briefly report on the impact of a low-reheating temperature on inferences of $\Lambda$CDM parameters and of the sum of neutrino masses from CMB and BAO data. To this purpose, we perform parameter estimation runs using the Planck+lensing+BOSS/eBOSS dataset combination, imposing prior cuts on the reheating temperature $\Trh < \{5,\,6,\,7,\, 10\}\,\MeV$. These values correspond to $\Neff<\{2.889,\,2.980,\,3.017,\, 3.042\}$. 
    In these runs, we let the six $\Lambda$CDM base parameters, the sum of neutrino masses  $\sum m_\nu$ and the reheating temperature vary at the same time. In this section we focus on constraints obtained using a flat $\Neff$ prior.

    The results are summarized in fig.~\ref{fig:params_panel}, where we plot the 1-dimensional posteriors for selected base and derived parameters. We also show, as a reference, vertical lines and shaded regions corresponding to parameter estimates and $68\%$ confidence interval in $\Lambda$CDM+$\sum m_\nu$ from the Planck TTTEEE+lensing+BAO dataset in~\cite{Planck:2018vyg}, as taken from table 6.21 of the parameters table available at the Planck legacy archive~\footnote{\url{https://pla.esac.esa.int/}}\footnote{Note that the results in the Planck 2018 parameters paper use a different BAO datasets; difference should however be minor, for the purpose of this section.}. We observe mild shifts in $\omega_c$, $H_0$ and, to a lesser extent, $n_s$. These shifts are likely to be caused by the low value $\Neff$ associated to a low reheating temperature. As discussed in one of the previous sections, this is the leading (albeit not the only) effect on the CMB spectra of a low $\Trh$. The direction of all the shifts that we observe in the figure do indeed match those of degeneracies between the parameters considered, and $\Neff$. 
    
    Building on this analysis, we further investigate the behavior of the limits on the sum of neutrino masses, $\sum m_\nu$, within low reheating scenarios. For this purpose, we performed runs using both $T_\mathrm{RH}$ and $N_\mathrm{eff}$ sampling strategies, applying progressively tighter priors to force the MCMC chains to explore regions of the parameter space beyond those that essentially reproduce the standard $\Lambda\mathrm{CDM} + \sum m_\nu$ scenario. We have considered the full Planck+lensing+BAO dataset combination, as well as Planck+lensing and Planck only datasets.

\begin{figure*}[t]
        \centering
        \includegraphics[scale=0.9]{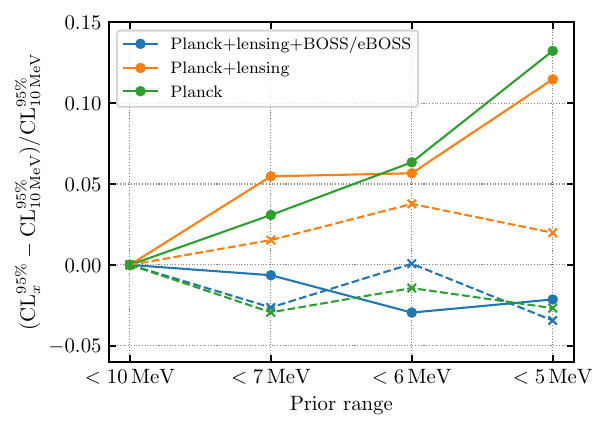}
        \caption{Relative variation of $95\%$ CL on the sum of neutrino masses, $\sum m_\nu$, for $T_\mathrm{RH}$ sampling (solid lines) and $N_\mathrm{eff}$ sampling (dashed lines). Different colors refer to different combinations of data sets.}
        \label{fig:combined_masses}
    \end{figure*}

    Our results from these runs are shown in \cref{fig:combined_masses}, where we plot the fractional difference of the 95\% upper bound with respect to the reference case $T_\mathrm{RH}<10\,\MeV$. For the $T_\mathrm{RH}$ sampling, we observe a slight, progressive relaxing of the neutrino mass constraints (up to nearly 15\%) with decreasing reheating temperature for the datasets that do not include BAO information. In particular, the Planck only result is consistent with our earlier findings in~\cite{deSalas:2015glj}. The trend is however not observed when the BAO data are included, or when the $N_\mathrm{eff}$ sampling is used. We can conclude that the neutrino mass bounds are in general very stable with respect to the reheating temperatures, for values of the latter within the allowed range.

\section{Summary of reheating temperature constraints}

    \begin{figure*}
        \centering
        \includegraphics[scale=.28]{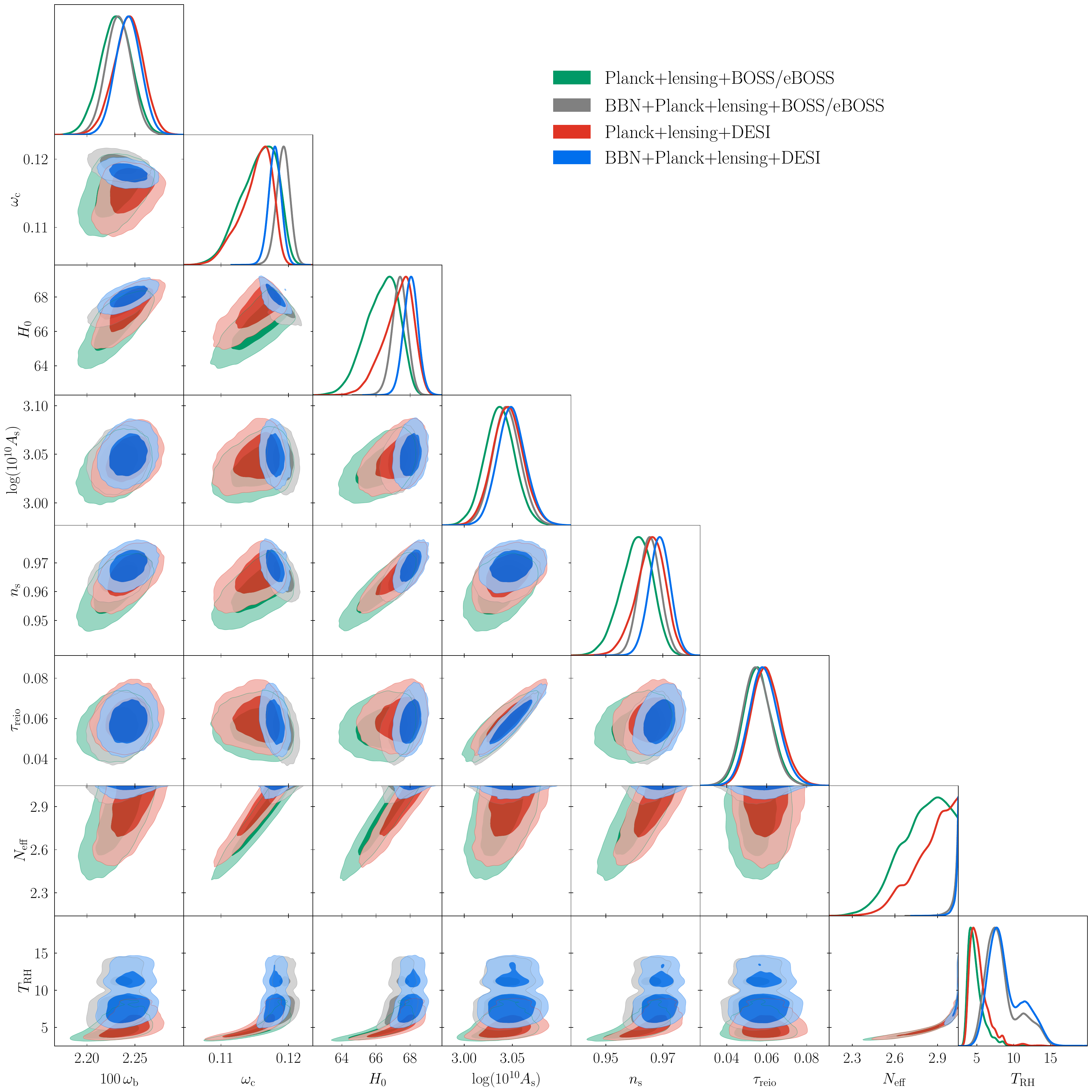}
        \caption{Triangle plot including one--dimensional posteriors and two-dimensional $68\%$ and $95\%$ credible regions for six cosmological parameters of the $\Lambda \mathrm{CDM}+T_\mathrm{RH}$ model, with $N_\mathrm{eff}$ sampling strategy, obtained from runs on all the data sets considered in this work. Numerical results with error bars are shown in \cref{tab:neff_cosmology}.}
        \label{fig:neff_cosmology}
    \end{figure*}

    \begin{figure*}
        \centering
        \includegraphics[scale=.33]{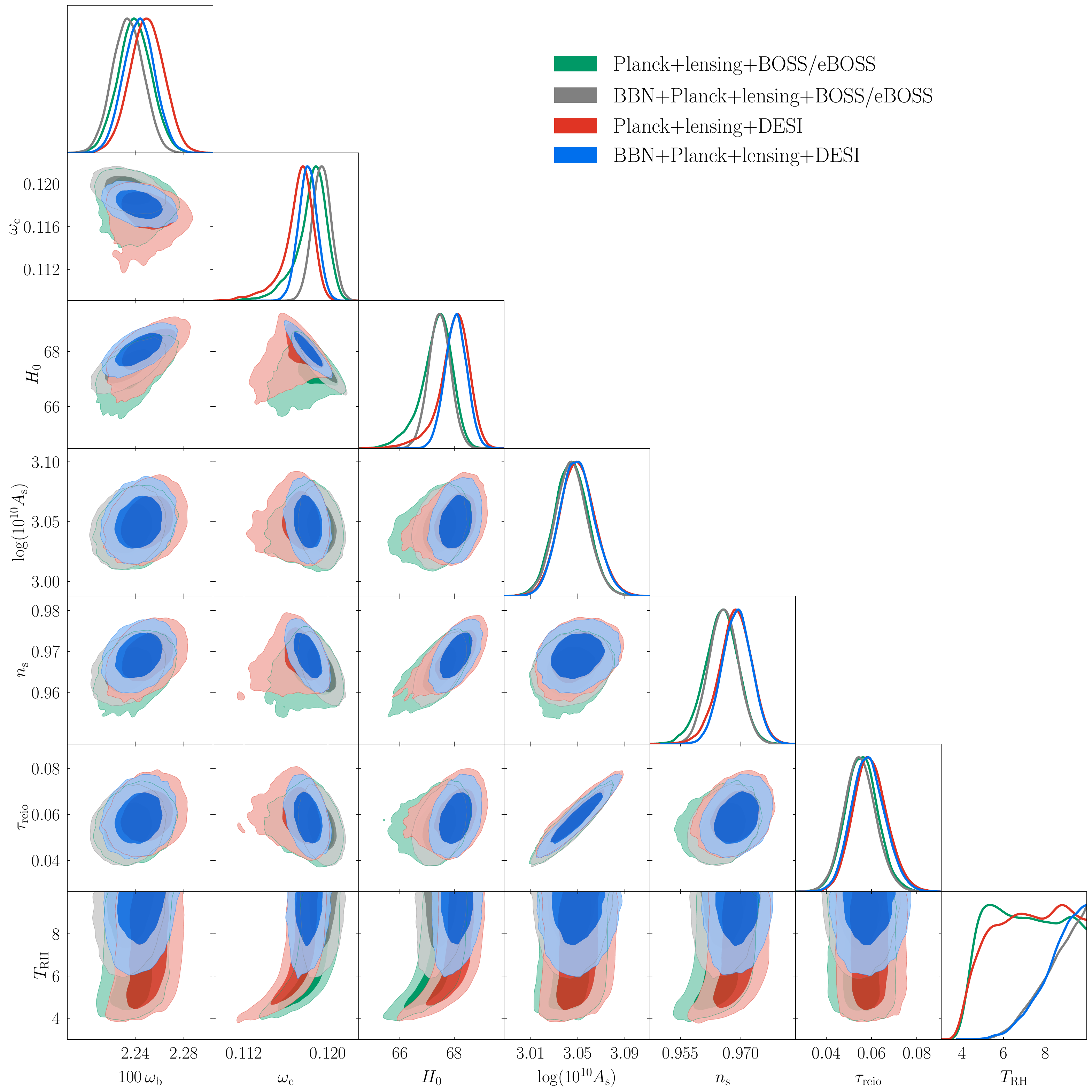}
        \caption{Triangle plot including one--dimensional posteriors and two-dimensional $68\%$ and $95\%$ credible regions for six cosmological parameters of the $\Lambda \mathrm{CDM}+T_\mathrm{RH}$ model, with $T_\mathrm{RH}$ sampling strategy, obtained from runs on all the data sets considered in this work. Numerical results with error bars are shown in \cref{tab:trh_cosmology}.}
        \label{fig:trh_cosmology}
    \end{figure*}

    \begin{figure*}
        \centering
        \includegraphics[scale=.33]{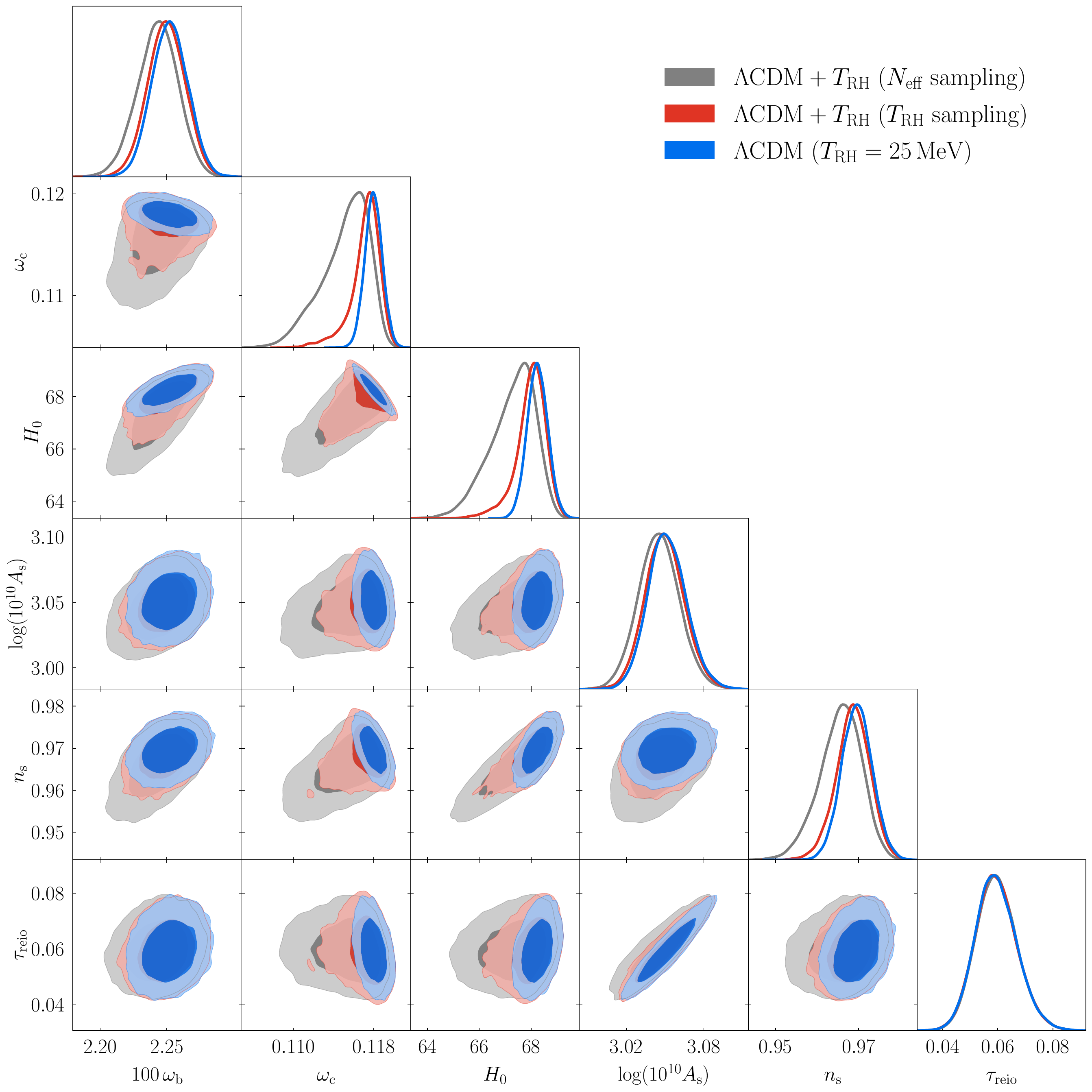}
        \caption{Triangle plot including one--dimensional posteriors and two-dimensional $68\%$ and $95\%$ credible regions for six cosmological parameters of the $\Lambda \mathrm{CDM}+T_\mathrm{RH}$ model (plus the standard $\Lambda\mathrm{CDM}$ model shown as a benchmark reference) obtained from runs with the Planck+lensing+DESI dataset. Note in particular the impact of the different sampling strategies on the posteriors of the various parameters. Numerical results with error bars are shown in \cref{tab:base_cosmology_DESI}.}
        \label{fig:sampling_cosmo}
    \end{figure*}

    We conclude summarizing here all the different constraint that we obtained using different datasets and sampling strategies detailed in the main text.  In addition to the DESI constraints reported in the main text, the entire analysis has been also conducted with the complementary ``Planck+lensing+BOSS/eBOSS'' dataset, where DESI measurements were substituted by 6dFGS/SDSS/BOSS/eBOSS. Specifically, we used low redshift galaxy samples from 6dFGS \cite{Beutler:2011hx} and SDSS-MGS \cite{Ross:2014qpa}, intermediate redshift emission line galaxies from eBOSS DR16 \cite{eBOSS:2020qek, eBOSS:2020fvk} and high redshift quasars \cite{eBOSS:2020lta, eBOSS:2020hur, eBOSS:2020gbb, eBOSS:2020uxp}, Lyman-$\alpha$ measurements and their correlations \cite{eBOSS:2020tmo}, from a combination of BOSS DR12 \cite{BOSS:2016wmc} and eBOSS DR16 \cite{eBOSS:2020yzd}. 
    
    The trends observed are consistent across all cases: the inclusion of BBN measurements significantly enhances the constraining power of the datasets. This effect is particularly pronounced when sampling on $N_\mathrm{eff}$. Replacing BOSS/eBOSS BAO measurements with those from DESI leads to slightly tighter limits, reflecting the larger constraining power of the DESI data. Finally, the $N_\mathrm{eff}$ uniform prior yields slightly looser limits on $T_\mathrm{RH}$ with respect to the flat $T_\mathrm{RH}$ prior, for the reasons outlined above.  
    \begin{align}
         &\quad& \textrm{flat } N_\mathrm{eff} \text{ prior} \;\; &\quad& \text{flat } T_\mathrm{RH} \text{ prior} \;\; \notag \\ 
        \text{Planck+lensing+BOSS/eBOSS:} &\quad& T_\mathrm{RH} > 3.62 \; \text{MeV} \;, &\quad&  T_\mathrm{RH} > 4.52 \; \text{MeV} \; . \\ 
        \text{Planck+lensing+DESI:} &\quad& T_\mathrm{RH} > 3.79 \; \text{MeV} \;, &\quad& T_\mathrm{RH} > 4.50 \; \text{MeV} \; . \label{eq:PLB_constraints} \\
        \text{BBN+Planck+lensing+BOSS/eBOSS:} &\quad& T_\mathrm{RH} > 5.57 \; \text{MeV} \;, &\quad& T_\mathrm{RH} > 6.71 \; \text{MeV} \; . \\ 
        \text{BBN+Planck+lensing+DESI:} &\quad& \bm{ T_\mathrm{RH} > 5.96 \; \textbf{MeV} } \;, &\quad& T_\mathrm{RH} > 6.76 \; \text{MeV} \; .
    \end{align}
    The result highlighted in bold is what we consider the most accurate and robust constraint from this work for the reasons explained in the main text and in this supplemental material. 
    Additional constraints presented here are included both for comparative purposes and to provide complementary, independent information. More detailed information about the variation of cosmological parameter constraints can be found in \cref{tab:neff_cosmology,tab:trh_cosmology} and \cref{fig:neff_cosmology,fig:trh_cosmology}.

    \begin{table}[h]
        \renewcommand{\arraystretch}{1.2}
        \centering
        \begin{ruledtabular}
            \begin{tabular}{ l c c c c}
                & {\bfseries Planck+lensing} & {\bfseries BBN+Planck+lensing} & {\bfseries Planck+lensing} & {\bfseries BBN+Planck+lensing} \\
                & {\bfseries +BOSS/eBOSS} & {\bfseries +BOSS/eBOSS} & {\bfseries +DESI} & {\bfseries +DESI} \\
                \midrule
                {\boldmath $100\,\omega_\mathrm{b}$} & $2.230^{+0.031}_{-0.033}   $ & $2.233 \pm 0.026   $ & $2.243^{+0.029}_{-0.031}   $ & $2.243^{+0.026}_{-0.025}   $\\
                {\boldmath$\omega_\mathrm{c}$} & $0.1155^{+0.0046}_{-0.0053}$ & $0.1192^{+0.0019}_{-0.0021}$ & $0.1151^{+0.0039}_{-0.0049}$ & $0.1180^{+0.0018}_{-0.0019}$\\
                {\boldmath$100 \, \theta_\mathrm{s}$} & $1.0426^{+0.0011}_{-0.00098}$ & $1.04198^{+0.00059}_{-0.00057}$ & $1.0426^{+0.0010}_{-0.00089}$ & $1.04210 \pm 0.00057$\\
                {\boldmath$\log ( 10^{10}A_\mathrm{s} )$} & $3.037 \pm 0.031   $ & $3.044^{+0.029}_{-0.028}   $ & $3.045^{+0.032}_{-0.031}   $ & $3.049^{+0.030}_{-0.028}   $\\
                {\boldmath$n_\mathrm{s}   $} & $0.961 \pm 0.011   $ & $0.9654^{+0.0074}_{-0.0073}$ & $0.9656^{+0.0091}_{-0.011} $ & $0.9689 \pm 0.0073$\\
                {\boldmath$\tau_\mathrm{reio}$} & $0.055^{+0.015}_{-0.014}   $ & $0.055^{+0.015}_{-0.014}   $ & $0.060^{+0.016}_{-0.014}   $ & $0.059^{+0.015}_{-0.014}   $\\
                {\boldmath$N_\mathrm{eff} $} & $> 2.51                    $ & $> 2.95                    $ & $> 2.58                    $ & $> 2.98                    $\\
                {\boldmath$H_0            $} & $66.3^{+1.7}_{-1.9}        $ & $67.38^{+0.87}_{-0.93}     $ & $67.3^{+1.5}_{-1.8}        $ & $68.04^{+0.83}_{-0.85}     $\\
            \end{tabular}
        \end{ruledtabular}
        \caption{$95\%$ Bayesian credible intervals for the seven parameters of the $\Lambda \mathrm{CDM}+T_\mathrm{RH}$ model, with $N_\mathrm{eff}$ sampling strategy, obtained from runs on all the data sets considered in this work. The corresponding posterior distributions are shown in \cref{fig:neff_cosmology}}
        \label{tab:neff_cosmology}
    \end{table}

    \begin{table}[h]
        \renewcommand{\arraystretch}{1.2}
        \centering
        \begin{ruledtabular}
            \begin{tabular}{ l c c c c}
                & {\bfseries Planck+lensing} & {\bfseries BBN+Planck+lensing} & {\bfseries Planck+lensing} & {\bfseries BBN+Planck+lensing} \\
                & {\bfseries +BOSS/eBOSS} & {\bfseries +BOSS/eBOSS} & {\bfseries +DESI} & {\bfseries +DESI} \\
                \midrule
                {\boldmath$100 \, \omega_\mathrm{b}$} & $2.240 \pm 0.027   $ & $2.234 \pm 0.025  $ & $2.250^{+0.027}_{-0.028}   $ & $2.243 \pm 0.025   $\\
                {\boldmath$\omega_\mathrm{c}$} & $0.1184^{+0.0026}_{-0.0033}$ & $0.1194 \pm 0.0018$ & $0.1171^{+0.0027}_{-0.0033}$ & $0.1181\pm 0.0018$\\
                {\boldmath $100 \, \theta_\mathrm{s}$} & $1.04209^{+0.00071}_{-0.00069}$ & $1.04194^{+0.00055}_{-0.00056}$ & $1.04223^{+0.00071}_{-0.00069}$ & $1.04207^{+0.00056}_{-0.00055}$\\
                {\boldmath$\log ( 10^{10}A_\mathrm{s} )$} & $3.045^{+0.029}_{-0.028}   $ & $3.045 \pm 0.028   $ & $3.050^{+0.031}_{-0.029}   $ & $3.050^{+0.030}_{-0.028}   $\\
                {\boldmath$n_\mathrm{s}   $} & $0.9651^{+0.0079}_{-0.0087}$ & $0.9657^{+0.0073}_{-0.0071}$ & $0.9686^{+0.0077}_{-0.0085}$ & $0.9690 \pm 0.0072$\\
                {\boldmath$\tau_\mathrm{reio}$} & $0.056^{+0.015}_{-0.014}   $ & $0.055^{+0.015}_{-0.014}   $ & $0.059^{+0.016}_{-0.014}   $ & $0.059^{+0.015}_{-0.014}   $\\
                {\boldmath$T_\mathrm{RH}  $} & $> 4.52                    $ & $> 6.71                    $ & $> 4.50                    $ & $> 6.76                    $\\
                {\boldmath$H_0            $} & $67.3^{+1.1}_{-1.3}        $ & $67.46 \pm 0.80     $ & $68.0^{+1.1}_{-1.3}        $ & $68.06 \pm 0.79     $\\
            \end{tabular}
        \end{ruledtabular}
        \caption{$95\%$ Bayesian credible intervals for the seven parameters of the $\Lambda \mathrm{CDM}+T_\mathrm{RH}$ model, with $T_\mathrm{RH}$ sampling strategy, obtained from runs on all the data sets considered in this work. The corresponding posterior distributions are shown in \cref{fig:trh_cosmology}}
        \label{tab:trh_cosmology}
    \end{table}

    \begin{table}[h!]
        \renewcommand{\arraystretch}{1.2}
        \centering
        \begin{ruledtabular}
            \begin{tabular}{ l c c c}
                & {\boldmath $\Lambda\mathrm{CDM}+T_\mathrm{RH}$} & {\boldmath $\Lambda\mathrm{CDM}+T_\mathrm{RH}$} & \boldmath $\Lambda\mathrm{CDM}$ \\
                Parameter & ($N^{FP}_\mathrm{eff}$ sampling) & ($T_\mathrm{RH}$ sampling) & ($T_\mathrm{RH} = 25 \, \text{MeV}$) \\ 
                \midrule
                {\boldmath$100 \, \omega_\mathrm{b}$} & $2.243^{+0.029}_{-0.031}   $ & $2.250^{+0.027}_{-0.028}   $ & $2.252 \pm 0.026   $\\
                {\boldmath$\omega_\mathrm{c}$} & $0.1151^{+0.0039}_{-0.0049}$ & $0.1171^{+0.0027}_{-0.0033}$ & $0.1180 \pm 0.0017$\\
                {\boldmath$100 \, \theta_\mathrm{s}$} & $1.0426^{+0.0010}_{-0.00089}$ & $1.04223^{+0.00071}_{-0.00069}$ & $1.04206 \pm 0.00056$\\
                {\boldmath$\log (10^{10} A_\mathrm{s})$} & $3.045^{+0.032}_{-0.031}   $ & $3.050^{+0.031}_{-0.029}   $ & $3.052^{+0.031}_{-0.028}   $\\
                {\boldmath$n_\mathrm{s}   $} & $0.9656^{+0.0091}_{-0.011} $ & $0.9686^{+0.0077}_{-0.0085}$ & $0.9696^{+0.0074}_{-0.0072}$\\
                {\boldmath$\tau_\mathrm{reio}$} & $0.060^{+0.016}_{-0.014}   $ & $0.059^{+0.016}_{-0.014}   $ & $0.059^{+0.016}_{-0.014}   $\\
                {\boldmath$H_0            $} & $67.3^{+1.5}_{-1.8}        $ & $68.0^{+1.1}_{-1.3}        $ & $68.25 \pm 0.79    $\\
            \end{tabular}
        \end{ruledtabular}
        \caption{$95\%$ Bayesian credible intervals for six cosmological parameters of the $\Lambda \mathrm{CDM}+T_\mathrm{RH}$ model obtained from runs with the Planck+lensing+DESI dataset. Different columns refers to different sampling strategies, as indicated in the table, plus the standard $\Lambda\mathrm{CDM}$ model shown as a benchmark reference. The corresponding posterior distributions are shown in \cref{fig:sampling_cosmo}.}
        \label{tab:base_cosmology_DESI}
    \end{table}

    In \cref{fig:sampling_cosmo} we show posterior distributions for the $\Lambda\text{CDM}$ parameters obtained in $\Lambda\text{CDM}+T_\mathrm{RH}$ model with the two sampling strategies, for Planck+lensing+DESI. We also show posterior for the same dataset obtained in $\Lambda\text{CDM}$ as a reference. The corresponding bayesian credible intervals are shown in \cref{tab:base_cosmology_DESI}.

\section{Sensitivity of next-generation CMB experiments}

    \begin{figure*}[t]
        \centering
        \includegraphics[scale=0.9]{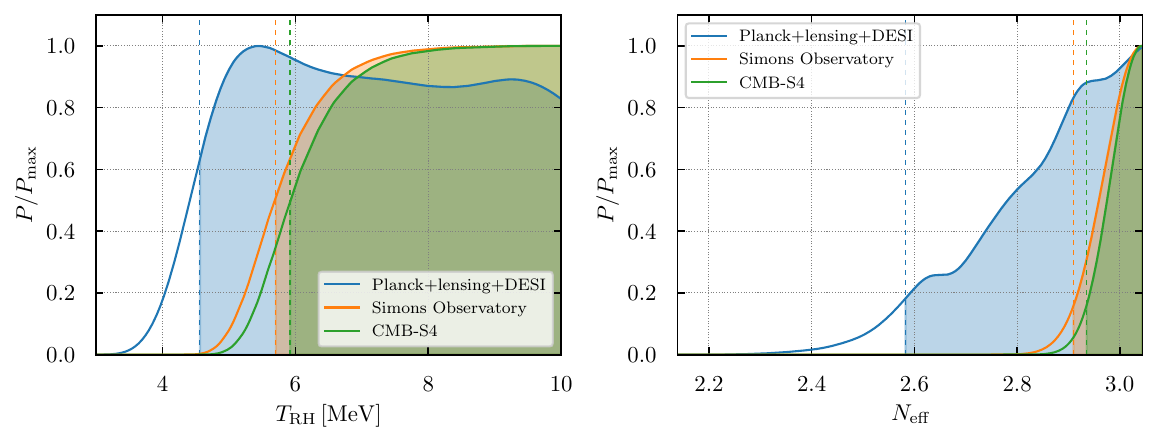}
        \caption{Posterior probability distributions for $T_\text{RH}$ (left) and $N_\text{eff}$ (right), referring to $T_\text{RH}$ and $N_\text{eff}$ sampling strategies, respectively. The curves represent forecasted experimental constraining power for Simons Observatory (orange), and CMB-S4 (green), compared to that of current data of Planck+lensing+DESI (blue). The shaded regions and the vertical dashed lines indicate the $68\%$ confidence intervals and lower limit, respectively.}
        \label{fig:forecasts}
    \end{figure*}

    Several CMB experiments will take data in the next years and we expect that they will improve with respect to current constraints on low-reheating scenarios. In this section we want to give a more quantitive assessment of their constraining power in this respect. In particular, we consider the reach of Simons Observatory (SO) \cite{Ade:2018sbj}, which is currently collecting data, and of a ``Stage-4'' CMB experiment providing a sub-percent precision measurement of $N_\text{eff}$. The latter is modeled over the baseline configuration of the CMB-S4 experiment described in Refs.~\cite{Abazajian:2019eic,CMB-S4:2020lpa}. We note that the CMB-S4 specifications are subject to change as they are currently being reevaluated within the collaboration, but target sensitivity to $\Neff$ is anyway expected to be similar. 
    
    A full MCMC forecast is beyond the illustrative scope of this section. Instead, in order to perform our forecasts we work under the assumption that we that, as long as future CMB observations are concerned, the sensitivity to $\Trh$ mainly comes through its effects on $\Neff$ and $Y_p$ - i.e., that the effect of neutrino spectral distortions on the evolution of perturbations is negligible at leading order (see above for a more detailed discussion).
    Given this assumption, we can exploit the information encoded in a marginalized parameter covariance matrix. For both SO and CMB-S4, we use the covariance matrices for $\omega_\mathrm{b}$, $N_\mathrm{eff}$ and $Y_p$ from \cite{Sabti:2019mhn,Escudero:2022okz}.
    These explicitly read
    \begin{center}
        \begin{tabular}{ccccc}
            \multicolumn{2}{c}{\textbf{Simons Observatory}} & & \multicolumn{2}{c}{\textbf{CMB-S4}} \\[4pt] 
            $\sigma \round{100 \omega_\mathrm{b}} = 0.0073 $ & $\rho \round{\omega_\mathrm{b}, Y_p} = 0.33$ & $\quad\qquad$ & $\sigma \round{100 \omega_\mathrm{b}} = 0.0047 $ & $\rho \round{\omega_\mathrm{b}, Y_p} = 0.22$ \\[2pt]
            $\sigma \round{N_\text{eff}} = 0.11$ & $\rho \round{\omega_\mathrm{b}, N_\text{eff}} = 0.072$ & $\quad\qquad$ & $\sigma \round{N_\text{eff}} = 0.081$ & $\rho \round{\omega_\mathrm{b}, N_\text{eff}} = 0.25$ \\[2pt]
            $\sigma \round{Y_p} = 0.0066$ & $\rho \round{N_\text{eff}, Y_p} = -0.86$ & $\quad\qquad$ & $\sigma \round{Y_p} = 0.0043$ & $\rho \round{N_\text{eff}, Y_p} = -0.84$
        \end{tabular}
    \end{center}
    where the $\sigma^2$s are the variances of the parameters and $\rho$ are their correlation coefficients.
    Given a covariance matrix, we write a multivariate gaussian likelihood in the parameters as 
    \begin{equation} \label{eq:future_like}
        \mathcal{L} \left( 
        \mathbf{\Theta}
        \right) \propto \exp \left[ - \frac{1}{2} \left( \mathbf{\Theta} - \mathbf{\Theta}_\text{fid} \right)^T \mathsf{{C}}^{-1} \left(\mathbf{\Theta} - \mathbf{\Theta}_\text{fid} \right) \right]
    \end{equation}
    where $\mathbf{\Theta} = \left( \omega_\mathrm{b}, N_\text{eff}, Y_p \right)$ is the vector of parameters, $\mathbf{\Theta}^\mathrm{fid}=\left(0.0224,\,3.044,\, 0.248\right)$ is their fiducial (observed) value, and $\mathsf{C}$ is the covariance matrix constructed as detailed above. Then, considering the mappings $\Neff(\Trh)$ and $Y_p(\Trh,\,\omega_b)$ we can rewrite \cref{eq:future_like} as a two-parameter likelihood 
    \begin{equation}
        \mathcal{L} \left( \omega_\mathrm{b}, T_\text{RH} \right) = \mathcal{L} \Big(
        \mathbf{\Theta}\left(\omega_\text{b}, \Trh \right) \Big) \; .
    \end{equation}
    Assuming separable priors, we the get as usual the posterior from Bayes' theorem as
    \begin{equation}
        P \left( \omega_\mathrm{b}, T_\text{RH} \right) \propto \mathcal{L} \left( \omega_\text{b}, T_\text{RH} \right) \Pi \left( \omega_\mathrm{b} \right) \Pi \left( T_\text{RH} \right) \; , 
    \end{equation} 
    which, after marginalizing over $\omega_\mathrm{b}$, gives the $1$-dim posterior $P \left(\Trh\right)$ from which we compute $95\%$ confidence intervals. Similarly to the analysis of current data presented in the main text, we consider both cases of a flat prior on $\Neff$ or $\Trh$.
    We find:
    \begin{align}
         &\quad& \text{flat } N_\mathrm{eff} \text{ prior} \;\; &\quad& \text{flat } T_\mathrm{RH} \text{ prior} \;\; \notag \\ 
        \text{Planck+lensing+DESI:} &\quad& T_\mathrm{RH} > 3.79 \; \text{MeV} \;, &\quad&  T_\mathrm{RH} > 4.50\; \text{MeV} \; . \\
        \text{Simons Observatory:} &\quad& T_\mathrm{RH} > 5.17 \; \text{MeV} \;, &\quad& T_\mathrm{RH} > 5.70 \; \text{MeV} \; . \\
        \text{CMB-S4:} &\quad& T_\mathrm{RH} > 5.41 \; \text{MeV} \;, &\quad& T_\mathrm{RH} > 5.91 \; \text{MeV} \; .
    \end{align}
    where we also reported the Planck+lesing+DESI results \ref{eq:PLB_constraints} for easier comparison.
    The full 1D posterior information is instead shown in \cref{fig:forecasts}.
     
    These results show how we can expect CMB observations alone to become, in a few years, competitive with the full combination of current datasets considered in our main analysis, that also includes information on primordial element abundances. This at least in part depends on the fact that the sensitivity of next-generation CMB observations to the helium abundance is comparable to the one of current astrophysical probes. Further inclusion of future measurements lensing, BAO and light element abundances will be able to increase this constraining power even more, possibly reaching the sensitivity to rule out very low reheating models with strong statistical significance.

\end{document}